\documentclass[sn-mathphys]{sn-jnl}

\usepackage{booktabs}
\usepackage{subcaption}
\usepackage[percent]{overpic}
\usepackage{pifont}
\usepackage{xfrac}
\usepackage{bm}
\usepackage{microtype}

\newcommand{\cmark}{\ding{51}}%
\newcommand{\xmark}{\ding{55}}%

\DeclareMathOperator*{\argmin}{arg\,min}

\newcommand{\ols}[1]{\mskip.5\thinmuskip\overline{\mskip-.5\thinmuskip {#1} \mskip-.5\thinmuskip}\mskip.5\thinmuskip} 
\newcommand{\olsi}[1]{\,\overline{\!{#1}}} 
\makeatletter
\newcommand\closure[1]{
	\tctestifnum{\count@stringtoks{#1}>1} 
	{\ols{#1}} 
	{\olsi{#1}} 
}
\long\def\count@stringtoks#1{\tc@earg\count@toks{\string#1}}
\long\def\count@toks#1{\the\numexpr-1\count@@toks#1.\tc@endcnt}
\long\def\count@@toks#1#2\tc@endcnt{+1\tc@ifempty{#2}{\relax}{\count@@toks#2\tc@endcnt}}
\def\tc@ifempty#1{\tc@testxifx{\expandafter\relax\detokenize{#1}\relax}}
\long\def\tc@earg#1#2{\expandafter#1\expandafter{#2}}
\long\def\tctestifnum#1{\tctestifcon{\ifnum#1\relax}}
\long\def\tctestifcon#1{#1\expandafter\tc@exfirst\else\expandafter\tc@exsecond\fi}
\long\def\tc@testxifx{\tc@earg\tctestifx}
\long\def\tctestifx#1{\tctestifcon{\ifx#1}}
\long\def\tc@exfirst#1#2{#1}
\long\def\tc@exsecond#1#2{#2}
\makeatother

\jyear{2021}%

\theoremstyle{thmstyleone}%
\theoremstyle{thmstyletwo}%

\theoremstyle{thmstylethree}%

\begin{document}

\title[Optimizing Mesh to Improve the TEA]{
	Optimizing Mesh to Improve the Triangular Expansion Algorithm for Computing Visibility Regions
}


\author*[1,2]{\fnm{Jan} \sur{Mikula}}\email{jan.mikula@cvut.cz}

\author[2]{\fnm{Miroslav} \sur{Kulich}}\email{miroslav.kulich@cvut.cz}


\affil[1]{\orgdiv{Department of Cybernetics}, \orgname{Faculty of Electrical Engineering, Czech Technical University in Prague}, \orgaddress{\street{Karlovo n\'{a}m\v{e}st\'{i} 13}, \city{Praha~2}, \postcode{12000}, \country{Czech Republic}}}

\affil[2]{\orgdiv{Czech Institute of Informatics, Robotics and Cybernetics}, \orgname{Czech Technical University in Prague}, \orgaddress{\street{Jugosl\'{a}vsk\'{y}ch partyz\'{a}n\r{u} 1580/3}, \city{Praha~6, Dejvice}, \postcode{16000}, \country{Czech Republic}}}



\abstract{
	This paper addresses the problem of improving the query performance of the \emph{triangular expansion algorithm} (TEA) for computing \emph{visibility regions} by finding the most advantageous instance of the \emph{triangular mesh}, the preprocessing structure.
	The TEA recursively traverses the mesh while keeping track of the visible region---the set of all points visible from a query point in a polygonal world. 
	We show that the measured query time is approximately proportional to the number of triangle edge expansions during the mesh traversal. 
	We propose a new type of triangular mesh that minimizes the expected number of expansions assuming the query points are drawn from a known probability distribution. 
	We design a heuristic method to approximate the mesh and evaluate the approach on many challenging instances that resemble real-world environments. 
	The~proposed mesh improves the mean query times by 12-16\% compared to the reference \emph{constrained Delaunay triangulation}. 
	The approach is suitable to boost offline applications that require computing millions of queries without addressing the preprocessing time.
	The implementation is publicly available to replicate our experiments and serve the community. 
}




\maketitle

\section{Introduction}\label{sec:intro}

Imagine a mobile robot with omnidirectional vision that is provided with a floor map of the building it is in and is given the task of finding a specific object with an unknown location somewhere on that floor. 
In order to be efficient, the robot must first plan a search path that optimizes for the earliest finding of the object and then follow that path. 
While constructing the search plan, the robot must be able to simulate its own vision at different locations in the map to ensure that particular significant areas and, eventually, the entirety of the map are covered. 
One way to implement the robot's vision in a known map is by using \emph{computational geometry} and the notion of \emph{visibility}. 

\begin{figure}
	\centering
	\begin{subfigure}[t]{0.3\textwidth}
		\begin{overpic}[width=\columnwidth]{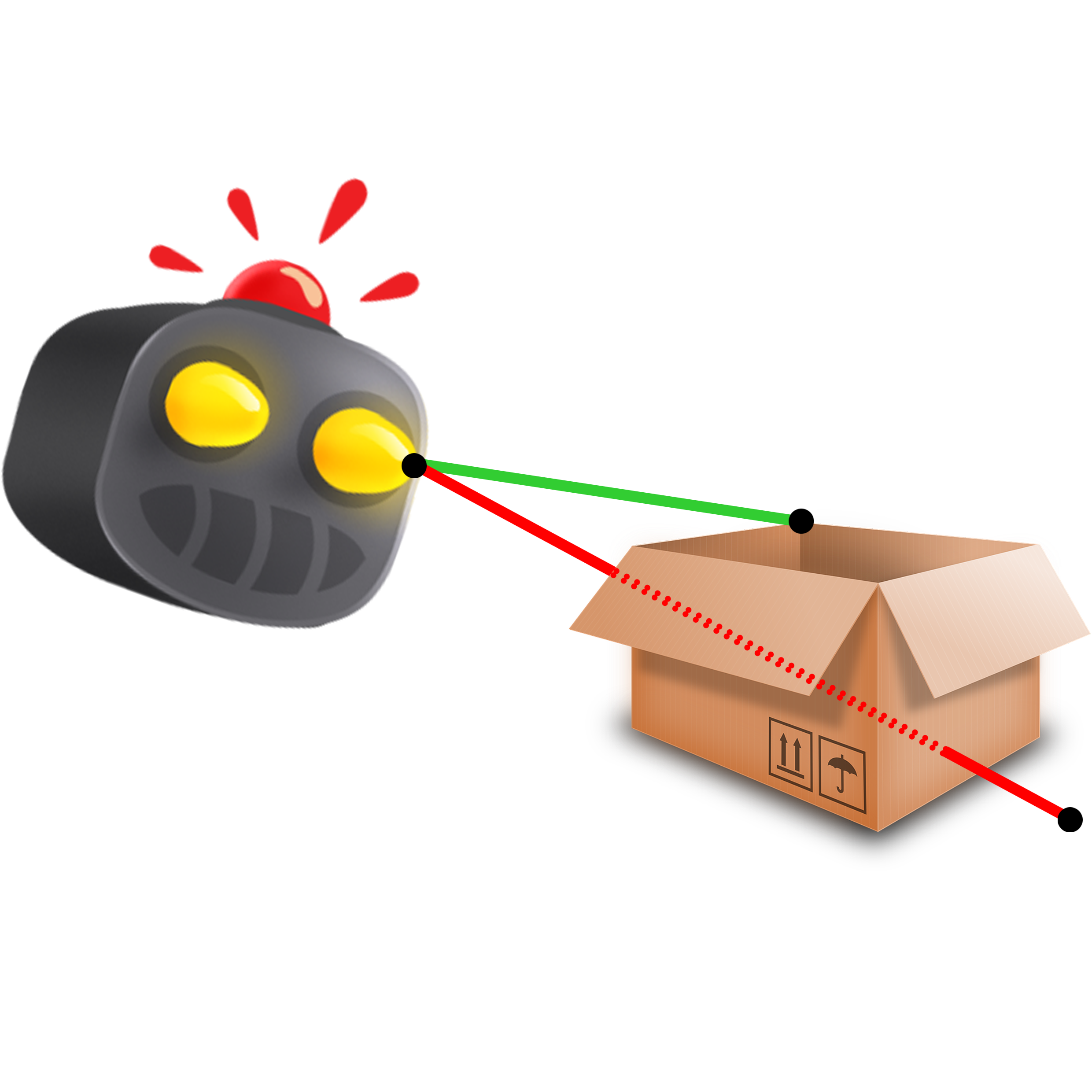}
			\put(39,62){$q$}
			\put(100,20){$p_{\text{\xmark}}$}
			\put(70,57){$p_{\text{\cmark}}$}
		\end{overpic}
		\caption*{Visibility concept.}
	\end{subfigure}
	\hfill
	\begin{subfigure}[t]{0.3\textwidth}
		\begin{overpic}[width=\columnwidth]{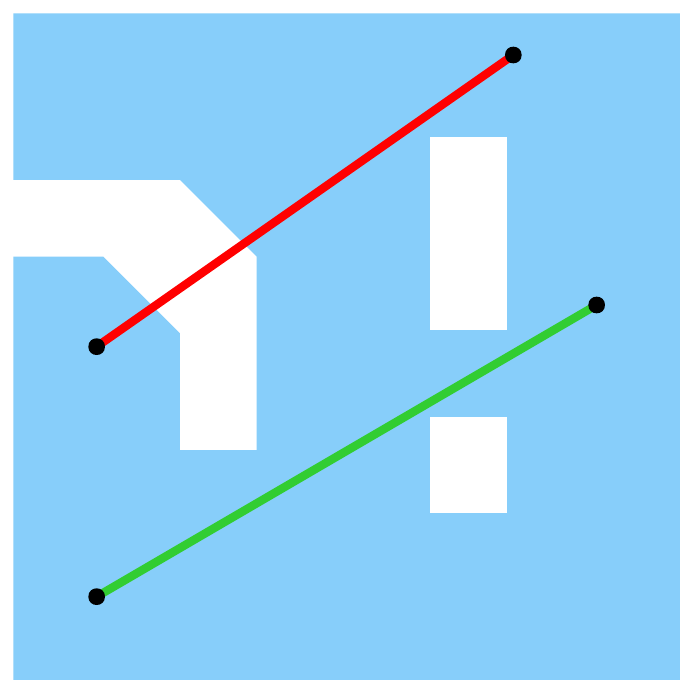}
			\put(3,87){$\mathcal{W}$}
			\put(10,41){$q_{\text{\xmark}}$}
			\put(70,82){$p_{\text{\xmark}}$}
			\put(10,5){$q_{\text{\cmark}}$}
			\put(80,46){$p_{\text{\cmark}}$}
		\end{overpic}
		\caption*{\hspace{-0.5em}\mbox{Visibility in polyg. domain.}}
	\end{subfigure}
	\hfill
	\begin{subfigure}[t]{0.3\textwidth}
		\begin{overpic}[width=\columnwidth]{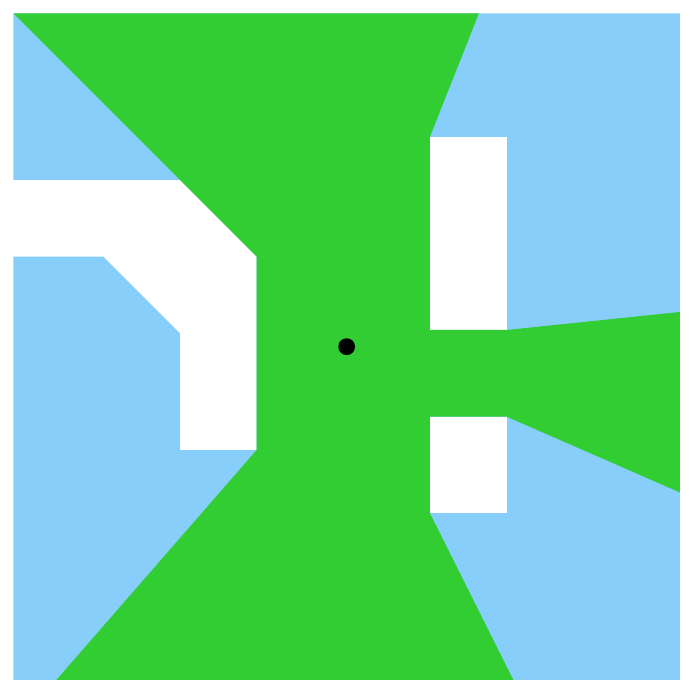}
			\put(3,50){$\mathcal{W}$}
			\put(46,41){$q$}
			\put(33,70){$\mathcal{V}(q)$}
		\end{overpic}
		\caption*{Visibility region.}
	\end{subfigure}
	\caption{Visibility and visibility region.}
	\label{fig:visibility}
\end{figure}
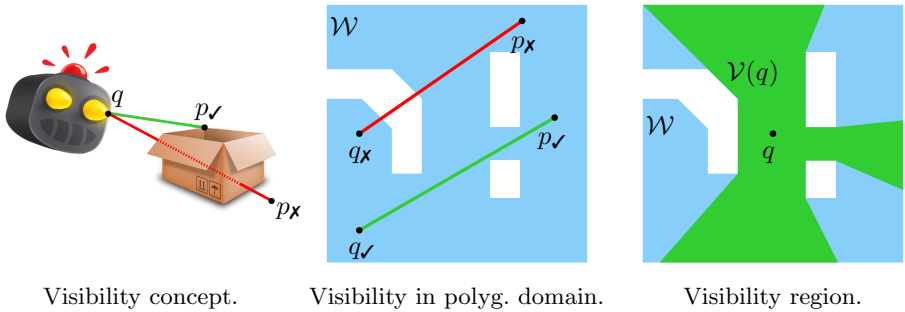

Visibility is a general concept to determine if an agent at known configuration $q$ in world $\mathcal{W}$ can see certain point $p$ of the world or if obstacles block its view. 
In \emph{polygonal domain}\footnote{
	$\mathcal{W} \subset \mathbb{R}^2$ is a polygon with holes that represents the interior of the map. 
	The outer boundaries and holes of $\mathcal{W}$ represent the obstacles that impose the motion and visibility constraints.
}, two points $q,p \in \mathcal{W}$ are visible to each other if segment $\closure{qp} \subset \mathcal{W}$.
In other words, the segment must be inside the polygon and must not properly intersect\footnote{
	The segment is allowed to touch the boundary but not cross it to the outside of $\mathcal{W}$.
} the polygon's boundary anywhere (including the holes).
If we want to simulate the entire field of view of the robot, we need to go further and determine all points visible from $q$, forming a connected subset of $\mathcal{W}$ known as the \emph{visibility region}~\cite{Davis1979}, formally defined as
\begin{equation}
	\mathcal{V}(q) = \big\{ v \in \mathbb{R}^2 \;\big\vert\; (\forall v)[\,\closure{vq} \subset \mathcal{W}\,] \big\}.
\end{equation}
Visibility regions take the form of star-shaped polygons that may have attached one-dimensional antennae, which are created when the observer is in line with two of the world's vertices, each blocking the view from the opposite side. 
In~practical use, the antennae are usually neglected; therefore, visibility regions are often regarded as simple polygons and called \emph{visibility polygons} accordingly. 
See the illustrations of all the basic terms in Fig.~\ref{fig:visibility}.

A visibility region can be computed by the \emph{rotational sweep algorithm} (RSA) \cite{Asano1985} in $\mathcal{O}(n\log n)$ time, where $n$ is the number of $\mathcal{W}$'s vertices.
However, in practice, the RSA is outstood by the \emph{triangular expansion algorithm}~(TEA)~\cite{Bungiu2014}, which runs in $\mathcal{O}(n^2)$ in the worst case but performs much better in practical scenarios. 
The authors of TEA demonstrate that their algorithm is two orders of magnitude faster than RSA in environments such as a cathedral or the interior of a country boundary. 
They also highlight that the worst-case time complexity only occurs in a class of carefully constructed instances, which are dissimilar to any real-world environment.

The TEA's idea is simple. 
As a \emph{preprocessing step}, a triangular mesh $\mathcal{T}$ is constructed to represent the polygonal world $\mathcal{W}$. 
The mesh is a collection of non-overlapping triangles, with vertices always being some of the vertices of $\mathcal{W}$, and their union is equal to $\mathcal{W}$.
The subsequent process of computing visibility regions for any number of input query points, the so-called \emph{query phase}, then utilizes only $\mathcal{T}$.
TEA's answer to query $q$ is initiated by locating triangle $\Delta_q \ni q$. 
Starting from $\Delta_q$, the neighboring triangles are recursively traversed while keeping track of the visibility region. 
TEA's first highlight is that it only traverses triangles visible from the query point. 
Thus, in some sense, it is output-sensitive, but not strictly so because it may traverse some triangles that do not contribute to the boundary of the result~\cite{Bungiu2014}. 
TEA's second highlight is that the operations used when traversing the mesh are extremely simple, essentially reducing to answering two orientation predicates per triangle traversal and computing at most two ray-segment intersections when a boundary edge is encountered.

\begin{figure}
	\centering
	\begin{subfigure}[b]{.19\linewidth}
		\includegraphics[width=\textwidth]{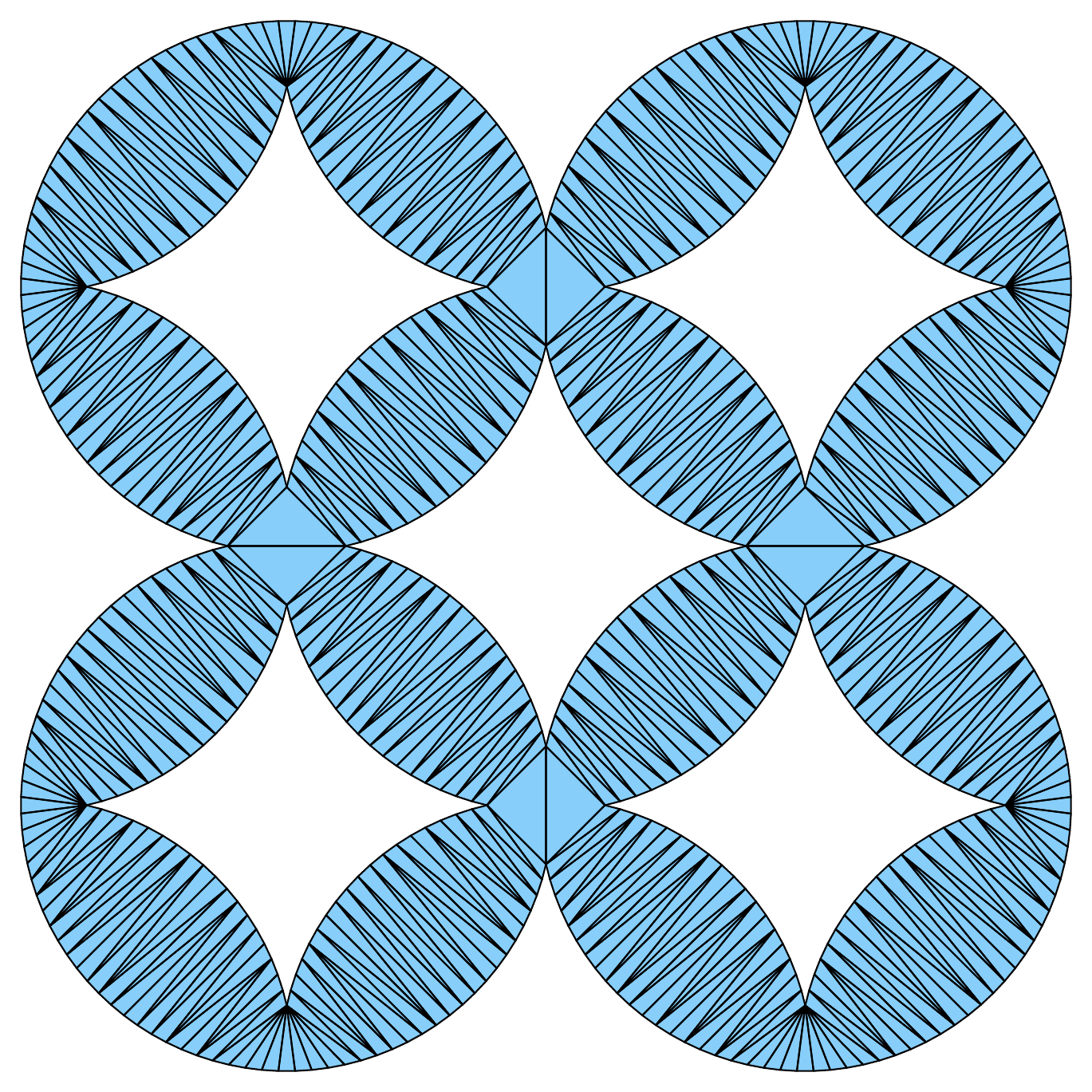}
		\caption{
			\centering CDT
		}
	\end{subfigure}
	\hfill
	\begin{subfigure}[b]{.19\linewidth}
		\includegraphics[width=\textwidth]{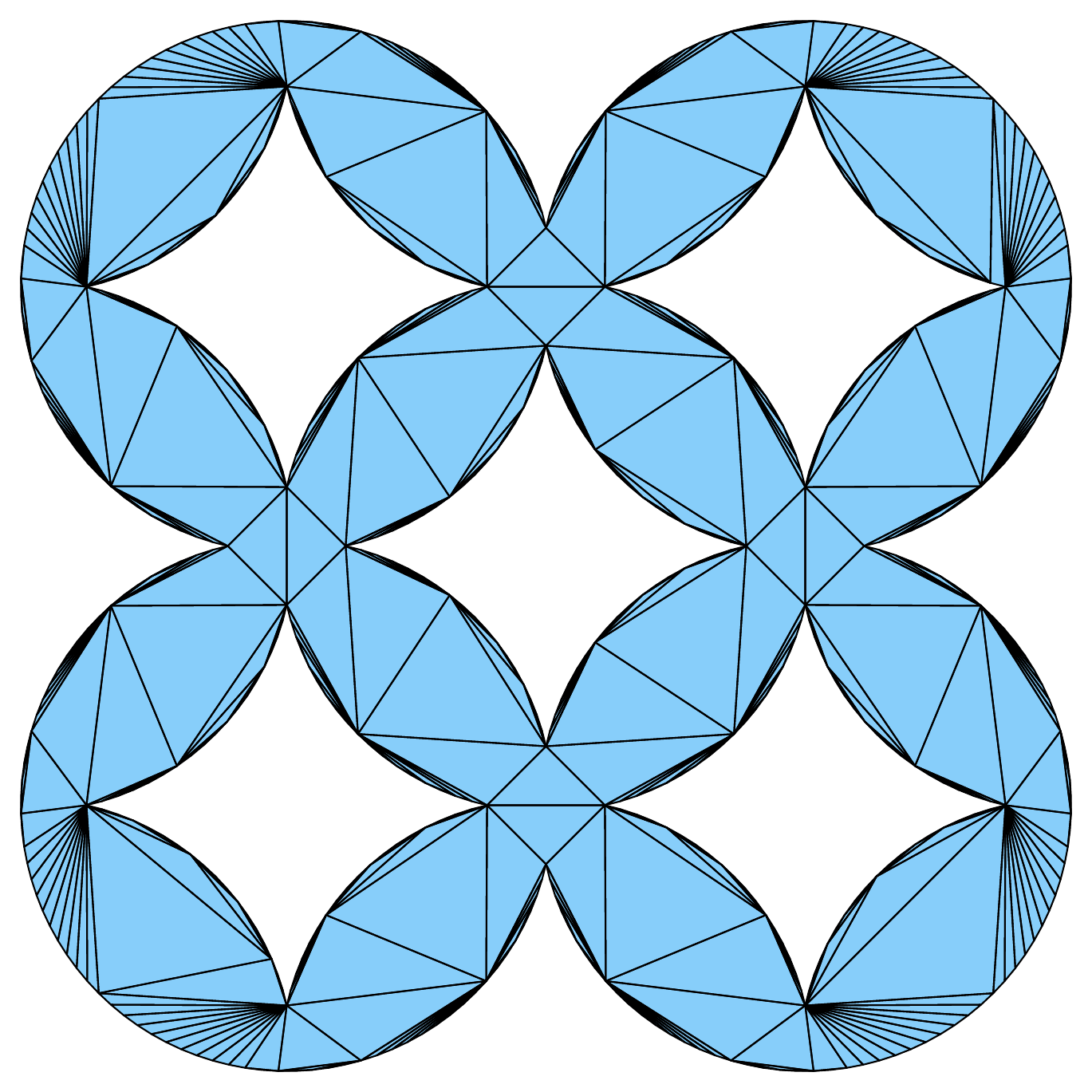}
		\caption{
			\centering MinVT
		}
	\end{subfigure}
	\hfill
	\begin{subfigure}[b]{.19\linewidth}
		\includegraphics[width=\textwidth]{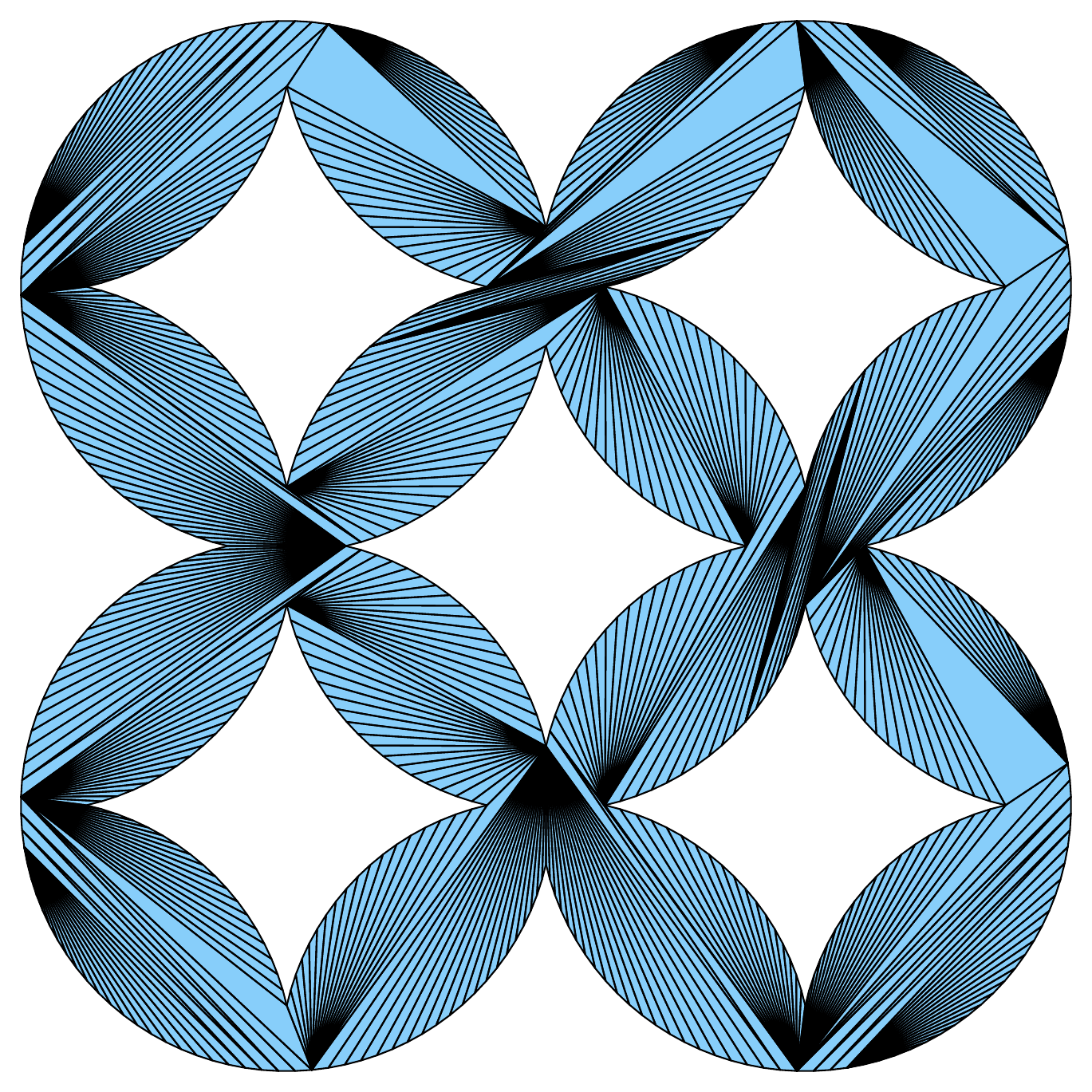}
		\caption{
			\centering MaxVT
		}
	\end{subfigure}
	\hfill
	\begin{subfigure}[b]{.19\linewidth}
		\includegraphics[width=\textwidth]{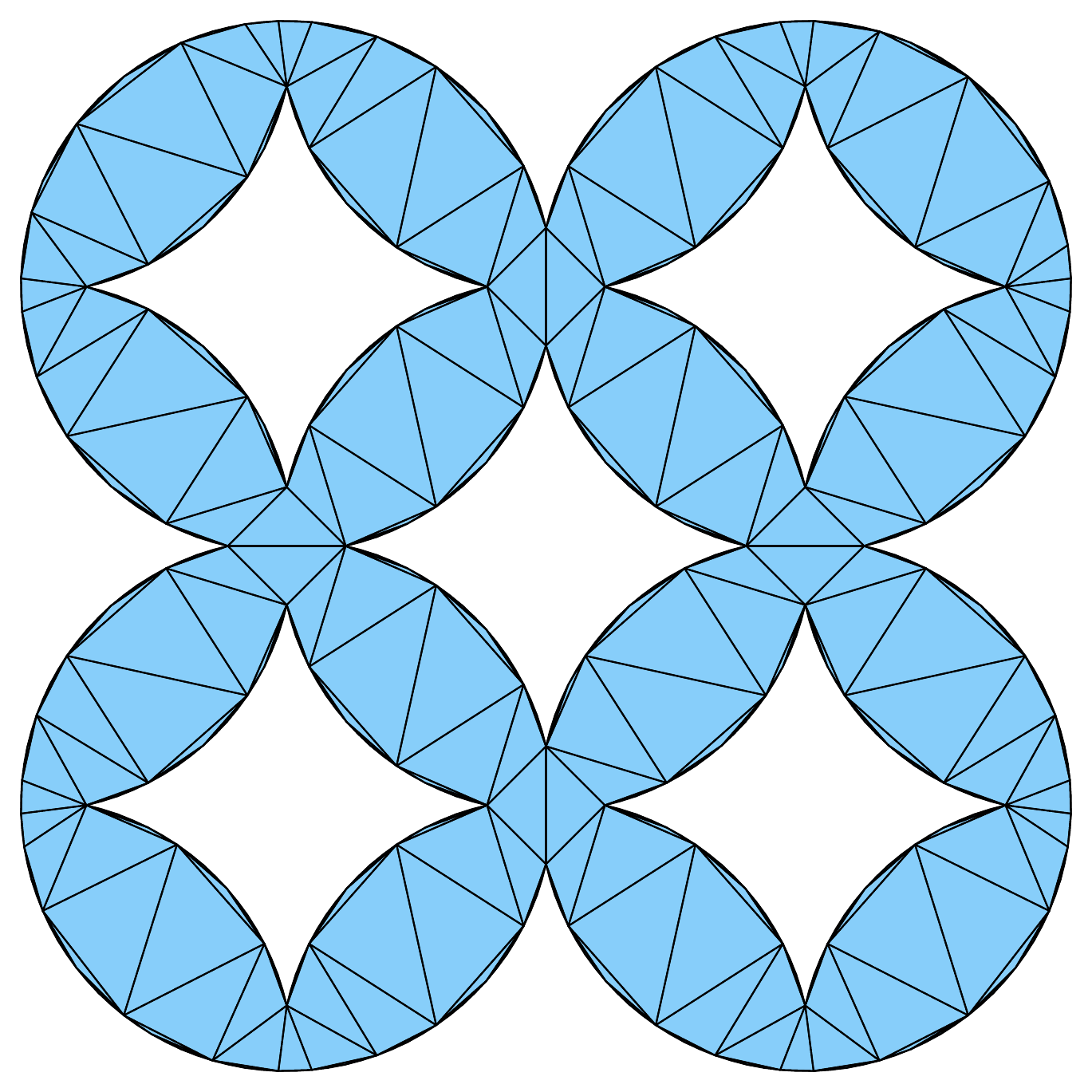}
		\caption{
			\centering MinLT
		}
	\end{subfigure}
	\hfill
	\begin{subfigure}[b]{.19\linewidth}
		\includegraphics[width=\textwidth]{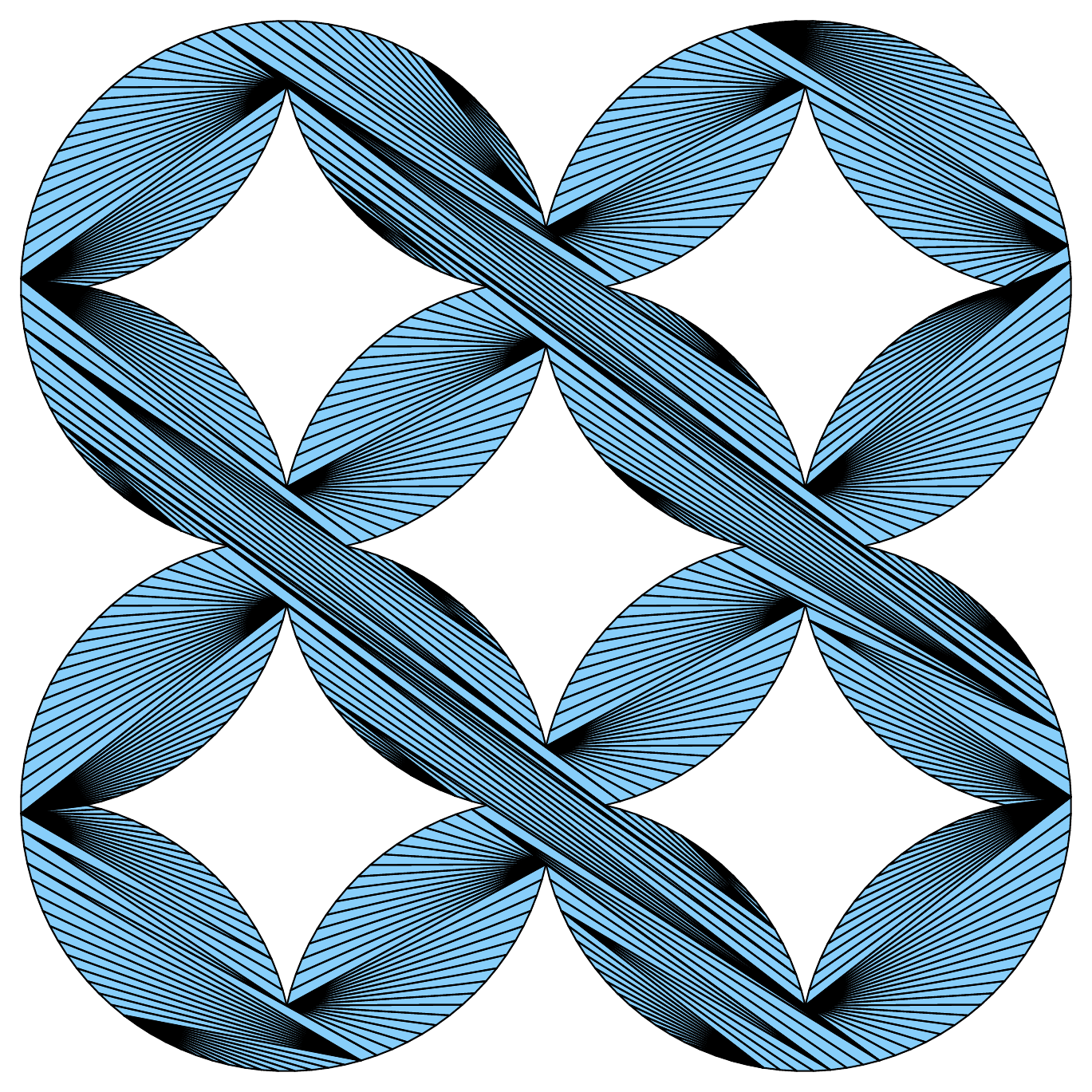}
		\caption{
			\centering MaxLT
		}
	\end{subfigure}
	\caption{Different triangular meshes of the same map (taken from~\cite{Mikula2022b}).}
	\label{fig:meshes}
\end{figure}

TEA's authors construct the triangular mesh by the \emph{constrained Delaunay triangulation}~(CDT)~\cite{Shewchuk2002}, likely due to its simplicity, fast run time, or availability of the implementation.
However, the question arises: is it the optimal choice? 
Does the mesh selection impact the query performance of the algorithm? 
If so, is there an optimal mesh, and how does CDT compare to other possible meshes? 
Interestingly, TEA's authors do not address any of these questions in~\cite{Bungiu2014}. 
Our earlier publication, as detailed in~\cite{Mikula2022b}, attempts to provide the answers.
We~show that when different triangular meshes are used (see Fig.~\ref{fig:meshes} for the examples), the~computational time required to answer the same queries on the same map may significantly differ. 
In~order to properly evaluate and compare the meshes, we~introduce the notion of the expected number of triangle edge expansions~$\eta_\mathcal{T}$, which strongly relates to the mean query performance of the algorithm~$\bar{t}_q$. 
We~propose a new type of mesh called MinVT, which approximately minimizes~$\eta_\mathcal{T}$ in polygons with holes, and design a heuristic method to construct it. 
Finally, we experimentally evaluate the CDT, MinVT, and other triangular mesh types in terms of $\eta_\mathcal{T}$ and $\bar{t}_q$ and show that MinVT indeed leads to the best metric values.

This paper builds upon~\cite{Mikula2022b} and introduces several contributions distinct from the previous work. 
We revisit the concepts underlying MinVT, reevaluating some of the assumptions made during the development of the optimal mesh. 
We also address and optimize the preprocessing time of our approach, i.e., the construction time of the MinVT mesh.
Moreover, the evaluation section is significantly expanded and now encompasses a diverse set of larger and more challenging instances taken from~\cite{Harabor2022}.
This paper represents a comprehensive analysis of the subject matter, contrasting with the original paper that serves as an introductory exploration of the topic.

Overall, the main contributions of this work are the following:
\begin{enumerate}
	\item We propose a new type of triangular mesh that minimizes the expected number of expansions performed by the TEA. 
	We argue why it should improve the algorithm's mean query time, mathematically formalize the idea, propose a heuristic algorithm to construct the mesh, transparently tune the algorithm's parameters, and experimentally evaluate the approach. 
	The evaluation shows that the approximation of the proposed mesh indeed provides the best mean query performance of the TEA compared to numerous other meshes, including the CDT. 
	\item We propose a simple modification of the TEA called d-TEA to cope efficiently with a more realistic visibility model with a limited visibility range~$d$. 
	Since the d-TEA expands a different number of edges than the TEA, we adapt the proposed mesh to consider this modification.
	The~evaluation shows that the adapted mesh consistently provides the lowest mean number of expansions. 
	However, oddly, for small values of $d$, the CDT performs the best in terms of mean query time instead.
	We provide a possible explanation together with a direction for future investigation. 
	\item We provide a simple, robust, and efficient open-source implementation of the TEA and the~proposed algorithms, along with the data and scripts used for our experiments, ensuring the replicability of our research. 
	The~implementation is a robust, versatile, and easy-to-use library designed to meet the diverse needs of the research community.
\end{enumerate}

The rest of this paper is organized as follows.
Sec.~\ref{sec:literature} reviews the related work with a subsection dedicated solely to the TEA~(\ref{sec:tea}).
Sec.~\ref{sec:problem} provides a general mathematical formalization of the optimal mesh~(\ref{sec:problem-general}) and then works in two simplifying assumptions~(\ref{sec:problem-assumptions}).  
Sec.~\ref{sec:solution} delivers the algorithms to construct the proposed mesh~(\ref{sec:MWT}-\ref{sec:MinVT}), introduces the d-TEA adaptation~(\ref{sec:d}) and details the implementation~(\ref{sec:imp}). 
Sec.~\ref{sec:results} evaluates our assumptions~(\ref{sec:eval-assumptions}), tunes our algorithms~(\ref{sec:tuning}), and presents the final evaluation results~(\ref{sec:eval-tea}-\ref{sec:eval-dtea}). 
Section~\ref{sec:conclusion} concludes the paper.
And finally, Appx.~\ref{sec:maps} provides some additional information about the benchmark instances.

\section{Related Work}
\label{sec:literature}

Computing visibility regions has been of high interest in computational geometry since 1979~\cite{Davis1979}.
Joe and Simpson~\cite{Joe1987} first provided a correct $\mathcal{O}(n)$ solution for computing visibility regions in simple polygons.
Heffernan and Mitchell~\cite{Heffernan1995} later proposed the optimal $\mathcal{O}(n + h\log h)$ algorithm for polygon with holes, where $h$ is the number of holes, improving over the previously best $\mathcal{O}(n\log n)$-time RSA by Asano~\cite{Asano1985}.
In the following decades many algorithms for a polygon with holes were published that provide tradeoffs between the preprocessing time and query time, e.g.: Zarei and Ghodsi~\cite{Zarei2008} with $\mathcal{O}(n^3\log n)$ preprocessing and $\mathcal{O}(K + \min(h,K)\log n)$ query time; Inkulu and Kapoor~\cite{Inkulu2009} with $\mathcal{O}(n^2\log n)$ preprocessing and $\mathcal{O}(K\log^2n)$ query time; and Chen and Wang~\cite{Chen2015} with $\mathcal{O}(n^2\log n)$ preprocessing and $\mathcal{O}(K + \log^2n+h\log(n/h))$ query time.

The research discussed thus far is theoretical, and many of the published algorithms are too complex for practical implementation.
In~contrast, our background is in the autonomous planning for long-term horizon missions with mobile robots, where we utilize visibility regions in complex real-world environments~\cite{Kulich14icra,Kulich17cor,Kulich19sensors}. 
In our field, visibility regions are useful when one needs to guard an art gallery with a set of static observers~\cite{ORourke1987}, establish a security guard’s route from which the whole boundary of an area is visible~\cite{Ntafos1992}, or~provide a guarantee that a room is free of intruders~\cite{Guibas1997}.
In our recent paper~\cite{Mikula2022a}, we consider a scenario involving a circular robot with radius $r$ equipped with an omnidirectional sensor with limited visibility range $d$. 
The objective is to plan the shortest route enabling the robot to visually inspect the entire environment. 
To tackle this NP-hard problem, we propose a heuristic algorithm that leverages the computation of visibility regions.
Notably, these regions need to be computed for a substantial number of query points, reaching into the millions, especially for the most challenging instances of the problem.

In the aforementioned applications, the significance lies more in the availability of implementation and the measured computational time on relevant problem instances than in the theoretical properties of the algorithms.
Bungiu~et~al.~\cite{Bungiu2014} implement Joe and Simpson's algorithm for simple polygons~\cite{Joe1987}, and Asano's RSA for polygons with holes~\cite{Asano1985} for the CGAL~\cite{CGAL} visibility package. 
In addition to these existing algorithms, Bungiu~et~al. propose and straightforwardly implement the TEA. 
As mentioned in the introduction, the authors then demonstrate that the TEA is two orders of magnitude faster than the RSA in real-world scenarios. 
From a practical standpoint, this establishes TEA as the state-of-the-art algorithm for computing visibility regions.
Subsec.~\ref{sec:tea} reviews the TEA as proposed by the authors~\cite{Bungiu2014}. 

Driven by the practical focus of our research, we conceived an idea to enhance the TEA by optimizing its triangular mesh. 
Our previous paper~\cite{Mikula2022b} asks the relevant questions on the subject and provides partial answers. 
The~current paper expands on this topic, offering an in-depth study as outlined in the introduction. 
Moreover, it's important to acknowledge the existence of parallel improvement ideas that our current work does not delve into. 
This paper focuses specifically on the TEA and the category of triangular meshes, exploring avenues for enhancement within this particular class.
A concurrent idea, which is not exclusive to our approach, is to consider more general class of meshes, such as convex polygonal ones referred to as  \emph{navigation meshes} in the pathfinding literature.

\emph{Pathfinding,} the task of finding the shortest obstacle-free path between two query points in a known map, is closely related to the visibility problems in the polygonal domain~\cite{Arkin2016}.
The state-of-the-art geometric pathfinding algorithm Polyanya~\cite{Cui2017}, which runs on convex polygonal meshes, was even adapted in~\cite{Shen2020} to compute all map vertices visible from another map vertex. 
We~want to emphasize that the two presented ideas are not necessarily competing but can complement each other. 
In~other words, the same considerations that led us to create the MinVT for the TEA might also guide the construction of a polygonal version of the optimized mesh for algorithms like Polyanya~\cite{Cui2017}.  
Although a~triangular mesh can be transformed into a convex polygonal mesh by merging some of its triangles into convex polygons, some other necessary adaptations may not be straightforward, such as modifying the dynamic programming approach later described in Alg.~\ref{alg:mwt-simple} for general navigation meshes.
Hence, a study of such a range exceeds the scope of this individual paper.

\subsection{Triangular Expansion Algorithm}
\label{sec:tea}

In this section, we revisit the TEA proposed in~\cite{Bungiu2014}, assuming triangular mesh $\mathcal{T}$ and query point $q$ is given.  
The mesh contains information about neighboring triangles, their edges, and vertices.

\begin{figure}
	\centering
	\begin{subfigure}[b]{.19\linewidth}
		\includegraphics[width=\textwidth]{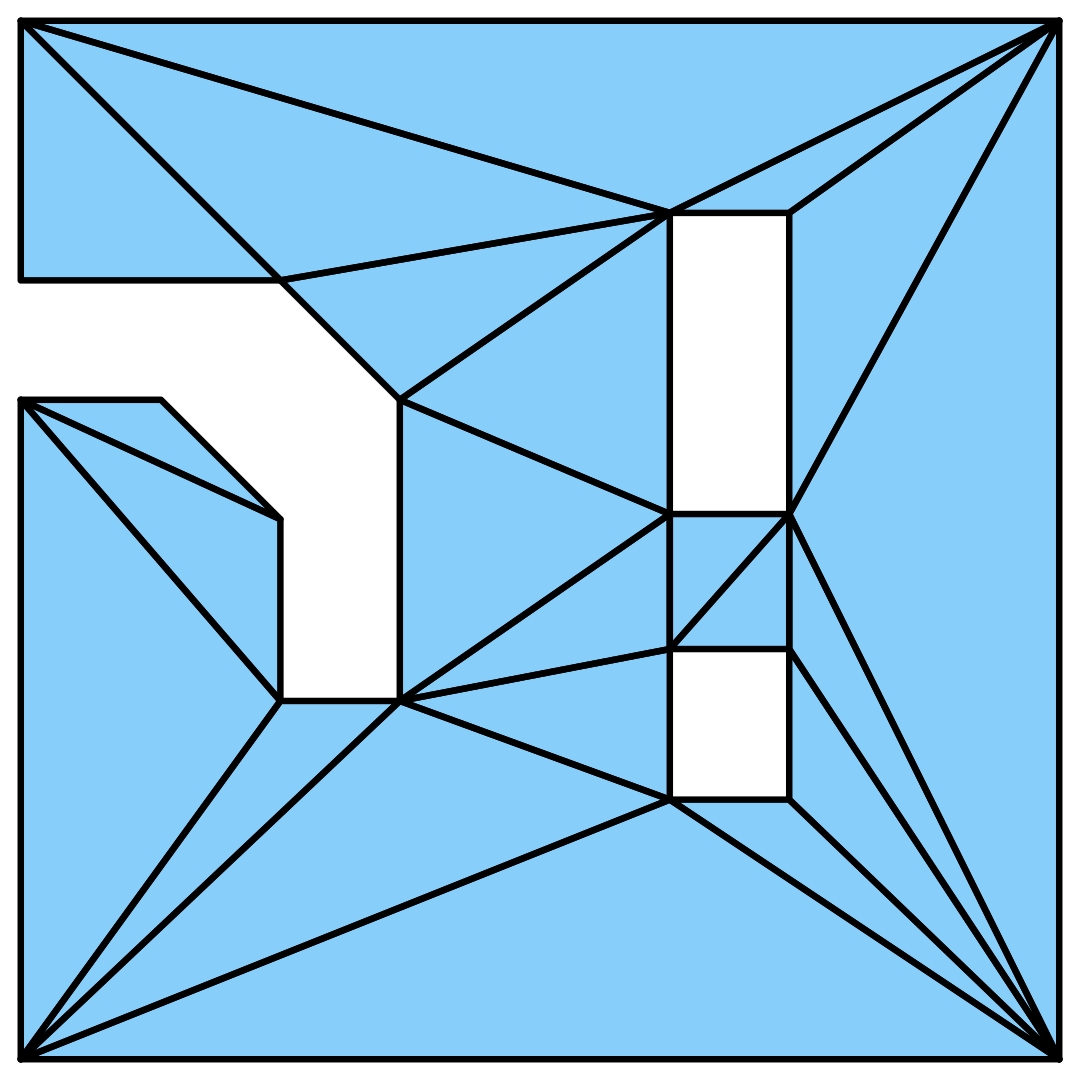}
		\caption*{Triang. mesh.}
	\end{subfigure}
	\hfill
	\begin{subfigure}[b]{.19\linewidth}
		\includegraphics[width=\textwidth]{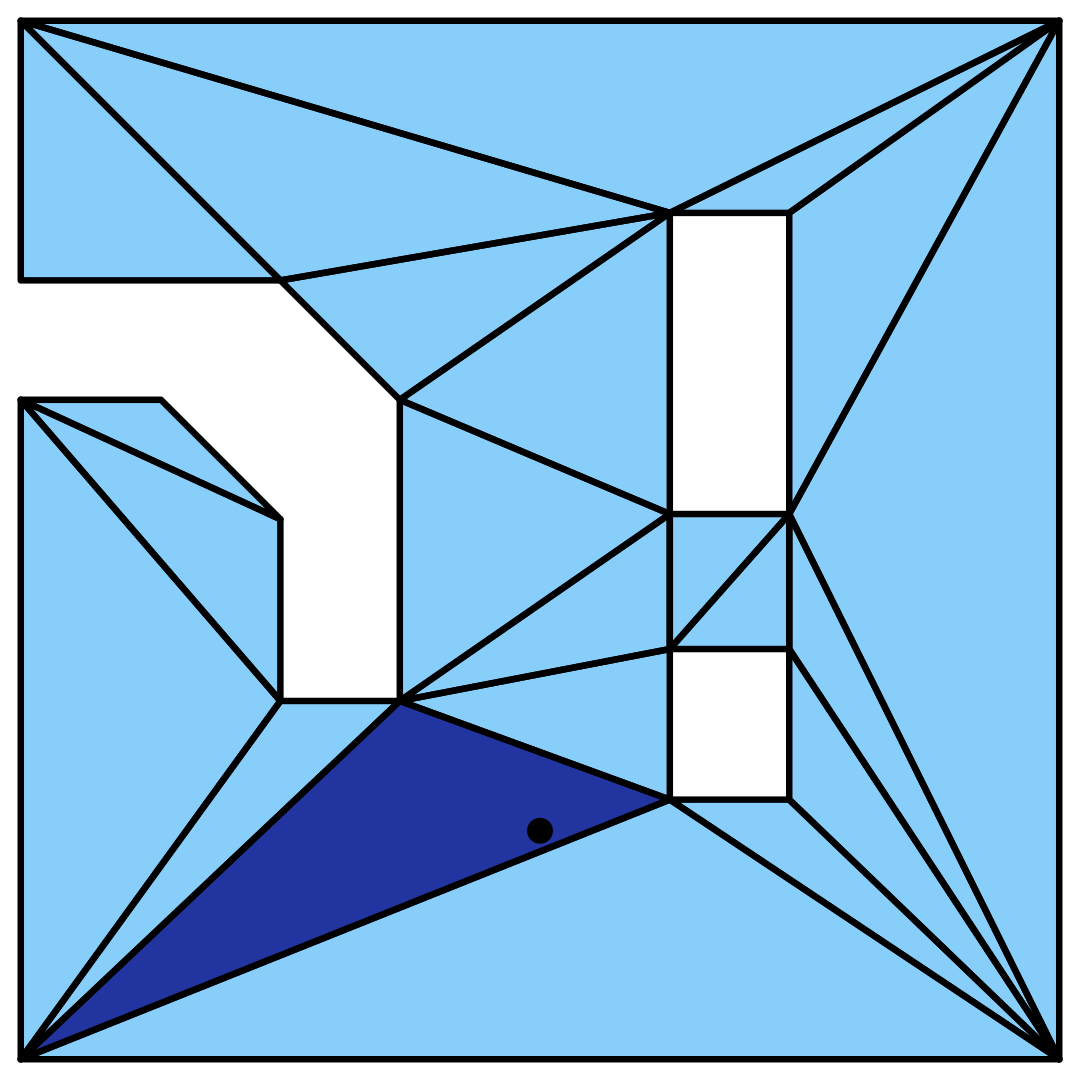}
		\caption*{Locating $\Delta_q$.}
	\end{subfigure}
	\hfill
	\begin{subfigure}[b]{.19\linewidth}
		\includegraphics[width=\textwidth]{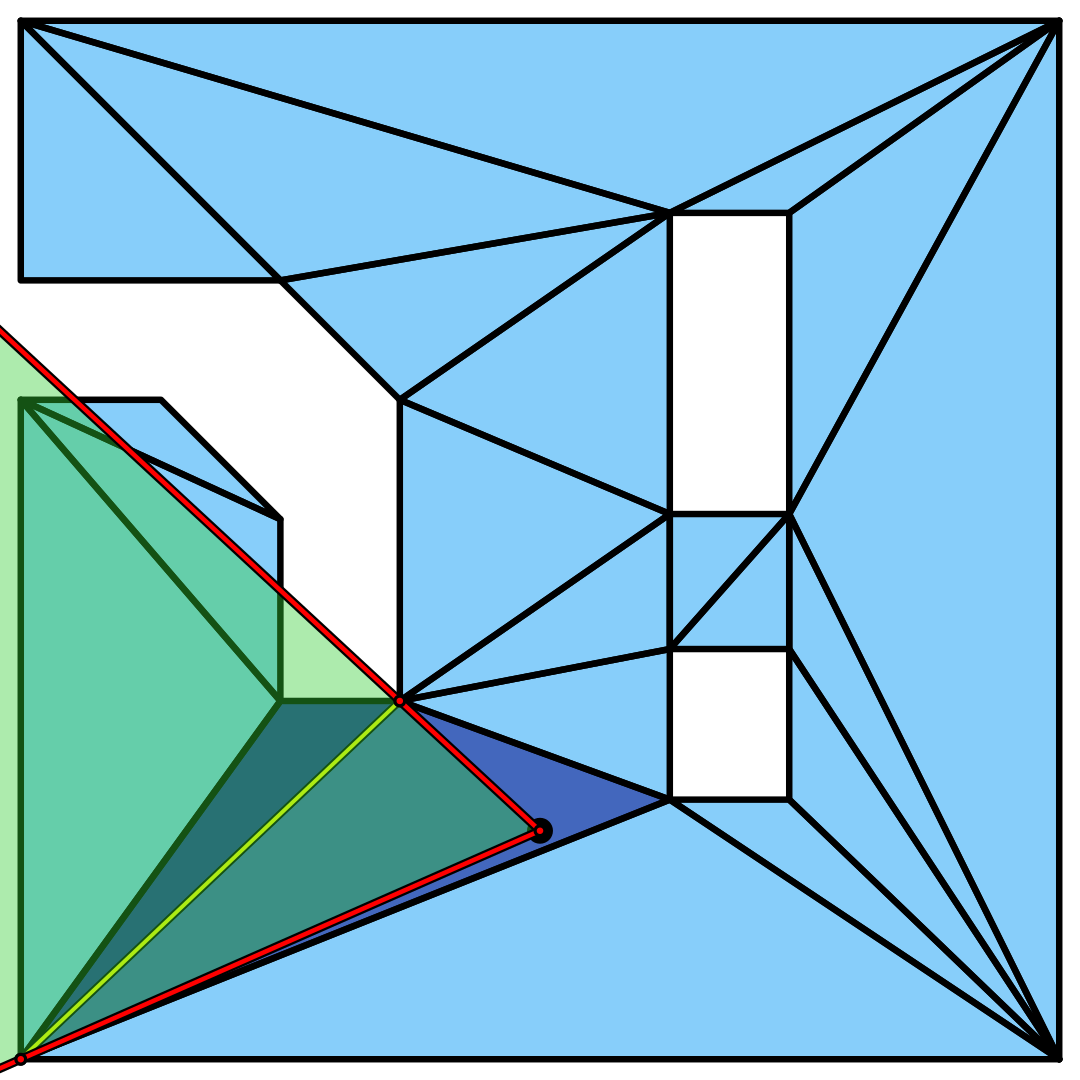}
		\caption*{\hspace{1em}\mbox{Traversing the triangulation.}}
	\end{subfigure}
	\hfill
	\begin{subfigure}[b]{.19\linewidth}
		\includegraphics[width=\textwidth]{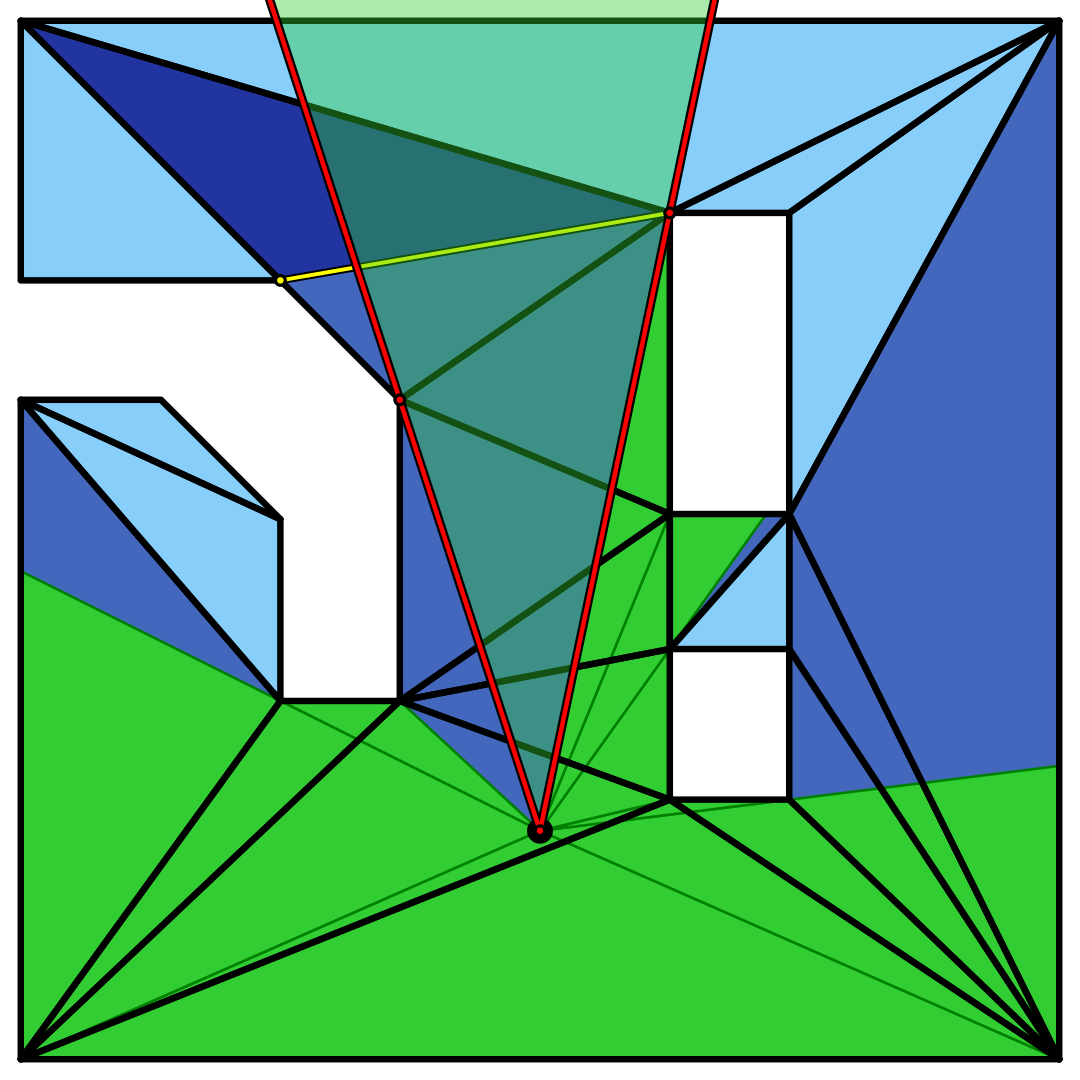}
		\caption*{}
	\end{subfigure}
	\hfill
	\begin{subfigure}[b]{.19\linewidth}
		\includegraphics[width=\textwidth]{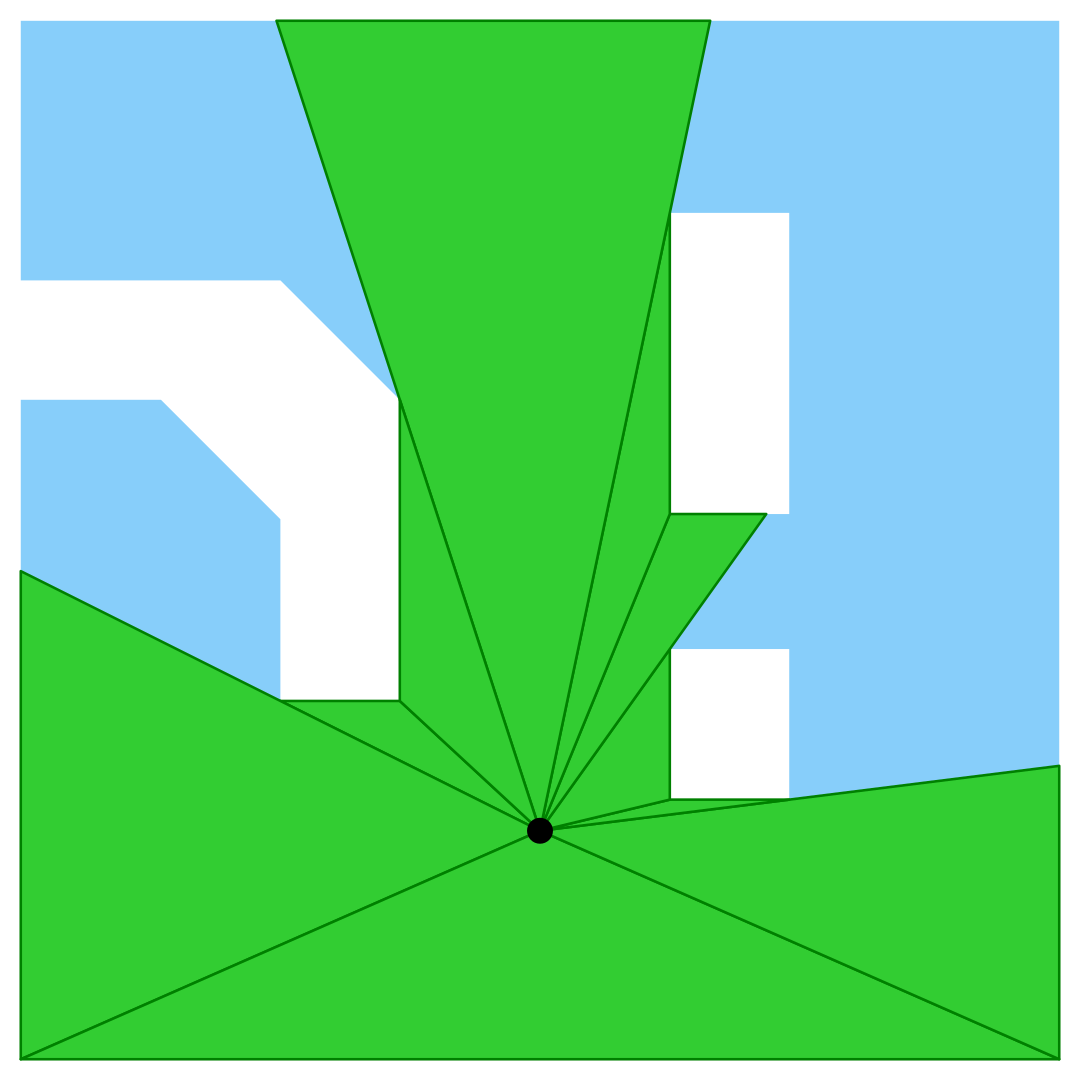}
		\caption*{$\mathcal{V}(q)$}
	\end{subfigure}
	\caption{The Triangular Expansion Algorithm.}
	\label{fig:TEA}
\end{figure}

The algorithm starts by locating triangle $\Delta_q$, $\Delta_q \in \mathcal{T}$ and $q \in \Delta_q$.
This~can be done by a simple walk in $\mathcal{O}(n)$, as the authors suggest.
When $\Delta_q$ is located, a recursive expansion procedure is then initiated for each of its edges. 
The~procedure expands along the current edge to the neighboring triangle while constraining the view of $q$ between the edge's endpoints. 
Within the new triangle, the two non-expanded edges are considered as candidates for the next recursion call only if they intersect the current view of $q$. 
The~recursion is invoked for any of those edges that neighbor another triangle.
Otherwise, the edge is identified as a \emph{boundary edge}, the current view of $q$ is ultimately constrained between its endpoints, stored, and the recursion does not propagate further from that point.
By convention, we do not consider the boundary edges to be expanded, even though the view reaches them.
Once no further expansions are possible, the resulting $\mathcal{V}(q)$ is the union of all the restricted views formed around $q$. 
Furthermore, an efficient TEA implementation takes advantage of preordering all the neighbor information in either clockwise (cw) or counterclockwise (ccw) order. 
The expansions then rotate around $q$ while adhering to the same order, and the union of the restricted views is naturally formed.
Visual illustrations of how the TEA operates can be found in Fig.~\ref{fig:TEA}.

According to the authors' analysis, the worst-case query time is $\mathcal{O}(n^2)$ because the recursion may split into two views $\mathcal{O}(n)$ times, and each view may reach $\mathcal{O}(n)$ triangles. 
However, splits into two views that reach the same triangle independently may occur only at $\mathcal{W}$'s holes. 
This implies that the worst-case query time is rather $\mathcal{O}(nh)$, where $h$ is the number of holes~\cite{Bungiu2014}.

\section{Problem Statement}\label{sec:problem}

\begin{table}[t]
	\begin{center}
		\begin{minipage}{\textwidth}
			\setlength{\tabcolsep}{0pt}
			\caption{A selection of the most frequent mathematical symbols.}\label{tab:math}%
			\begin{tabular*}{\textwidth}{@{\extracolsep{\fill}}ll@{\extracolsep{\fill}}}
				\toprule
				Symbols & Meaning \\
                \midrule
                $\mathcal{W},\mathrm{Bd}$ & Polygonal world $\mathcal{W} \subset \mathbb{R}^2$ with $n$ vertices and $h$ holes, and its boundary $\mathrm{Bd}(\mathcal{W})$. \\
                $q,p,\overline{qp}$ & Points $q,p \in \mathbb{R}^2$, and a line $\overline{qp}$ segment connecting them. \\
                $\mathcal{T},E,\Delta$ & Triangular mesh $\mathcal{T}$, its edges $E(\mathcal{T})$, and a triangle $\Delta \in \mathcal{T}$. \\ 
                $\mathcal{V}, \mathcal{V}_e$ & Visibility region from a point and from a segment. \\
                $t_q,\eta_q,\eta_q^h$ & Computational time, no. expansions, and no. expanded edges. \\
                $\bar{t}_q,\eta_\mathcal{T},\eta_\mathcal{T}^h$ & Avg. comp. time, expected no. expansions, expected no. expanded edges. 
                \\
				\botrule
			\end{tabular*}
		\end{minipage}
	\end{center}
\end{table}

Our objective is to minimize the average number of elementary computations, which, in turn, leads to minimizing edge expansions of the TEA. 
Drawing insight from TEA's authors, we understand that the algorithm selectively expands triangle edges visible from the query point and adjacent to two triangles. 
It's crucial to emphasize that edges neighboring only one triangle are part of $\mathcal{W}$'s boundary and must be incorporated into any valid mesh. 
These details are revisited in Sec.~\ref{sec:tea}. 
Given that only edges visible from the query point are expanded, it naturally follows that the mesh minimizing the expected visibility of its edges must be optimal. 
This constitutes the fundamental concept of our approach.
The remainder of this section is dedicated to formalizing this idea.
For the reader's convenience, Tab.~\ref{tab:math} presents the most frequently used mathematical symbols.

\subsection{Optimal Mesh Formalization}
\label{sec:problem-general}

We start by defining our quality metrics $t_q$ and $\eta_q$. Assuming $\mathcal{T}$ and $q$ are given: $t_q$ is the computational time to obtain $\mathcal{V}(q)$ using the TEA, and $\eta_q$ is the number of edge expansions recorded in the same scenario. 
The time $t_q$ is a measurable value that we aim to minimize, but it is influenced by hidden variables such as implementation, hardware setup, state of memory, and other concurrent processes.
On the other hand, the number of expansions $\eta_q$ depends solely on $\mathcal{T}$ and $q$. 
Given that the TEA's elementary computations scale linearly with the number of expansions, we can anticipate that $t_q$ will be proportional to $\eta_q$. 
This serves as an assumption, upon which we base our approach to minimize $\eta_q$ (its expected value, to be precise). 
Nevertheless, both metrics are utilized to evaluate our approach at the conclusion of our analysis.

We have established $\eta_q$ as the definitive quality metric for the pair $(q, \mathcal{T})$. 
However, to define the optimal mesh, we require a metric that is independent of any specific query point. 
To achieve this, we let the query point be described by the random variable $Q$ with a known two-dimensional \emph{probability density function}~(PDF)~$f_{X,Y}$ defined inside the polygonal world $\mathcal{W}$.
Then, the expected number of expansions given the triangular mesh $\mathcal{T}$ is defined as
\begin{equation}
	\eta_{\mathcal{T}}(\mathcal{T}) = \mathbb{EXP}\big[\eta_q(Q, \mathcal{T})\big] =  \iint_{\mathcal{W}}\eta_q\big((x, y), \mathcal{T}\big)\cdot f_{X,Y}(x, y)\,\mathrm{d}x\,\mathrm{d}y.
	\label{eq:etaT}
\end{equation}
Now we need to formally define $\eta_q$. 
We have already mentioned one way to compute it: initialize the TEA with $\mathcal{T}$ and compute $\mathcal{V}(q)$ while counting the number of expansions in the process. 
Note that this can be expressed in terms of how many times an edge is expanded:
\begin{equation}
	\eta_q(q, \mathcal{T}) = \sum_{e\,\in\,E(\mathcal{T})}\eta_e(e, q),
	\label{eq:etaQT}
\end{equation}
where $E(\mathcal{T})$ is the set of all $\mathcal{T}$'s edges, and $\eta_e(e, q)$ is the count of how many independent views of the TEA recursive expansion procedure reach~$e$ from~$q$.
For every view reaching $e$ from $q$, there must be a unique segment on $e$ whose all points are visible from $q$. 
We can enumerate the endpoints of all such segments using set-builder notation:
\begin{equation}
	\mathcal{S}(e, q)=\big\{(u, v)\in\mathbb{R}^2 \times \mathbb{R}^2\;\big\vert\; \closure{uv} \subset e\text{ and }(\forall p \in \closure{uv})[\,\closure{qp} \subset \mathcal{W}\,]\big\}.
	\label{eq:set}
\end{equation}
The expansion count $\eta_e$ can now be expressed using the cardinality of this set:
\begin{equation}
	\eta_e(e, q) = \begin{cases}
		0 & \text{if $e \subset \mathrm{Bd}(\mathcal{W})$,} \\
		\vert \mathcal{S}(e, q) \vert & \text{else,}
	\end{cases}
	\label{eq:etaE}
\end{equation}
where $\mathrm{Bd}(\mathcal{W})$ is the boundary of $\mathcal{W}$.
The uppercase in \eqref{eq:etaE} is to avoid counting the boundary edges, which, by convention, are considered as not expanded even if the view reaches them, as stated in Sec.~\ref{sec:tea}. 
By substituting \eqref{eq:etaE} into \eqref{eq:etaQT} and then replacing the result into \eqref{eq:etaT}, and utilizing the sum rule of integration, we obtain
\begin{align}
	\eta_\mathcal{T}(\mathcal{T}) &= \hspace{-1em} \sum_{e\,\in\,E_{\mathit{in}}(\mathcal{T})}\iint_{\mathcal{W}}\big\vert \mathcal{S}\big(e, (x, y)\big) \big\vert\cdot f_{X,Y}(x, y)\,\mathrm{d}x\,\mathrm{d}y,
	\label{eq:etaT2}
\end{align}
where $E_{\mathit{in}}(\mathcal{T}) = \{e\,\in\,E(\mathcal{T}) \;\vert\; e\,\not\subset\,\mathrm{Bd}(\mathcal{W})\}$ represents the set of $\mathcal{T}$'s interior edges (all non-boundary edges), and it follows from Eq.~\eqref{eq:etaE} (consider the two cases). 
Finally, let $\mathcal{T}_{\mathcal{W}}(\mathcal{W})$ be the set of all triangular meshes that represent polygon with holes~$\mathcal{W}$. 
Then, the optimal mesh is
\begin{equation}
	\mathcal{T}^\star = \argmin_{\mathcal{T} \,\in\, \mathcal{T}_{\mathcal{W}}(\mathcal{W})}\eta_{\mathcal{T}}(\mathcal{T}).
	\label{eq:Tstar}
\end{equation}
The formalization is done. 
Eq.~\eqref{eq:set},~\eqref{eq:etaT2}, and~\eqref{eq:Tstar} describe the optimal triangular mesh for the TEA without the notion of the algorithm itself.
Recall the two underlying assumptions: the number of expansions is proportional to the actual query time performance of the algorithm, and the PDF of the query points is known.

\subsection{Simplifying Assumptions}
\label{sec:problem-assumptions}

\begin{figure}
	\centering
	\begin{subfigure}[b]{.19\linewidth}
		\includegraphics[width=\textwidth]{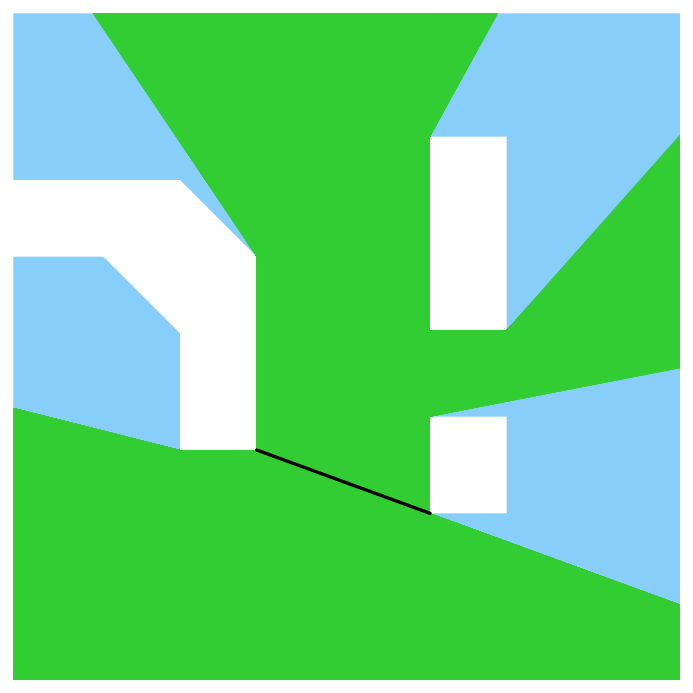}
	\end{subfigure}
	\hfill
	\begin{subfigure}[b]{.19\linewidth}
		\includegraphics[width=\textwidth]{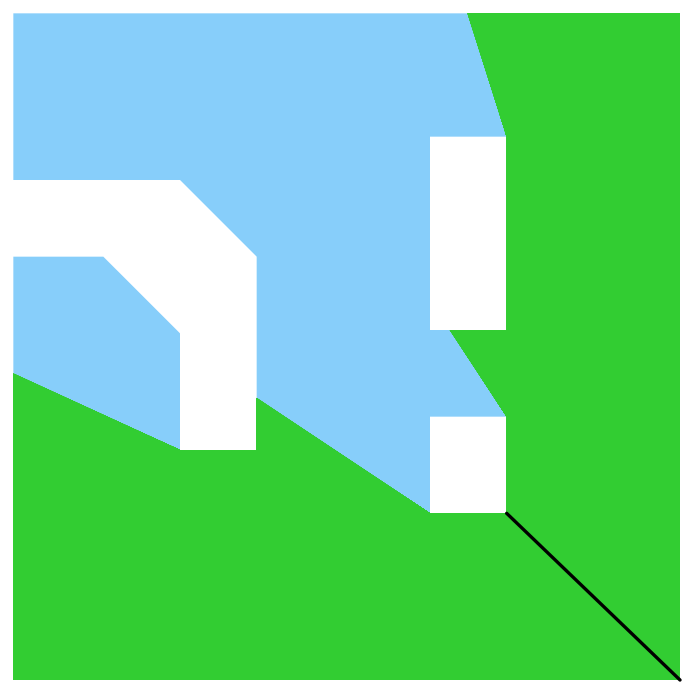}
	\end{subfigure}
	\hfill
	\begin{subfigure}[b]{.19\linewidth}
		\includegraphics[width=\textwidth]{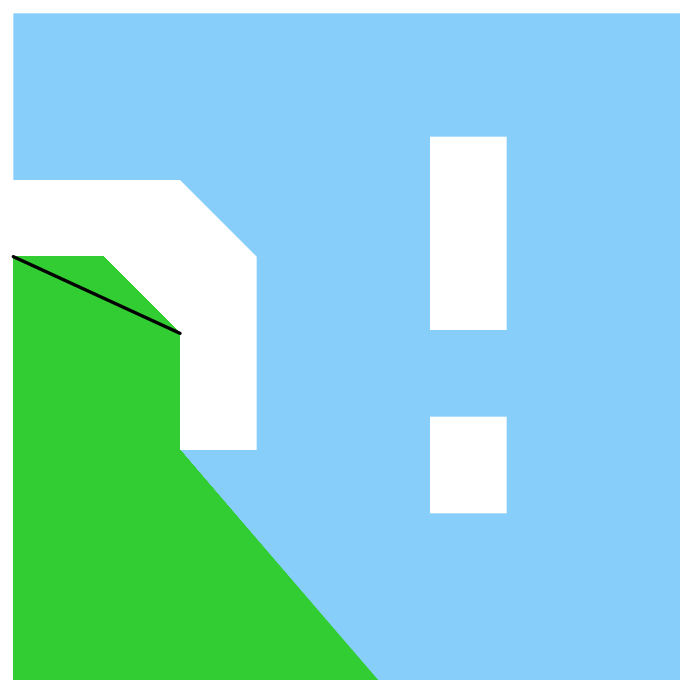}
	\end{subfigure}
	\hfill
	\begin{subfigure}[b]{.19\linewidth}
		\includegraphics[width=\textwidth]{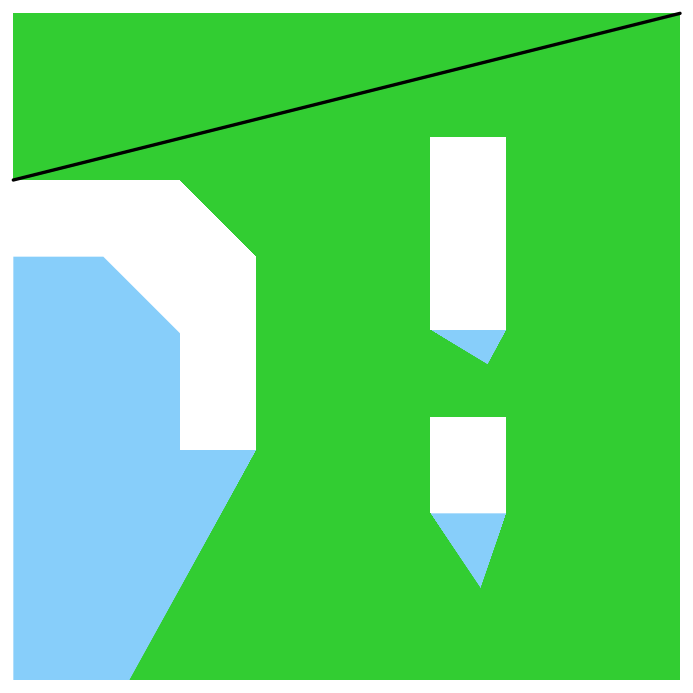}
	\end{subfigure}
	\hfill
	\begin{subfigure}[b]{.19\linewidth}
		\includegraphics[width=\textwidth]{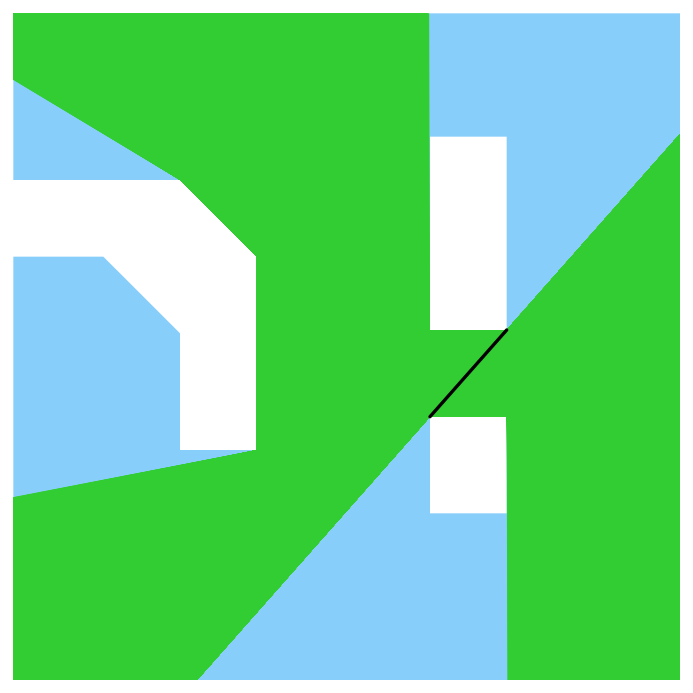}
	\end{subfigure}
	\caption{Examples of regions $\mathcal{V}_e(e)$ (green) visible from segments $e$ (black).}
	\label{fig:Ve}
\end{figure}

Although the formalism we introduced precisely describes what we want to achieve, it does not provide any practical way of constructing and evaluating the optimal mesh.
For that, we need to introduce two simplifying assumptions:
\begin{enumerate}
	\item\label{assumption1} 
	In many applications, we do not know anything about the probability distribution of a certain random variable. 
    In such a case, it is often useful to assume that all possible values are equally likely. 
    This assumption is well-known as the \emph{principle of indifference}~\cite{Keynes1921} in the literature. 
    \emph{When applied to the query points, it implies modeling them using a uniform probability density function (PDF) defined inside $\mathcal{W}$.}
    We use this assumption both in constructing $\mathcal{T}^\star$ and in the evaluation, where we draw query points from a uniform distribution using a pseudo-random generator.
	\item\label{assumption2} The second assumption can be expressed simply: \emph{No edge is expanded more than once.} 
    This means that $\mathcal{W}$ must be a simple polygon or have the property that it is impossible to have multiple independent views from one query point reach the same triangle edge. 
    In our formal language, the assumption can be written as:
	\begin{equation}
		\mathcal{W} \in \Big\{ \mathcal{W}' \;\big\vert\; (\forall q \in \mathcal{W}' )(\forall \mathcal{T}\in\mathcal{T}_{\mathcal{W}}(\mathcal{W}'))(\forall e\,\in\,E(\mathcal{T}))\big[\,\vert \mathcal{S}(e, q) \vert \leq 1\,\big]\Big\}.
		\label{eq:assumption}
	\end{equation}
	We use it only when constructing the mesh but do not enforce it in its evaluation. 
    This is the first step away from the exact solution towards a heuristic one. 
    We denote $\mathcal{T}^\star$ as $\mathcal{T}^{h\star}$ and $\eta$ as $\eta^h$ to emphasize that the assumption was used.
\end{enumerate}
Now we work these assumptions into the equations from the previous section.
Assumption~\ref{assumption2} implies that the number of expansions is the same as the number of expanded edges since no edge can be expanded more than once.
An edge is expanded if it is at least partially visible from $q$; therefore, Eq.~\eqref{eq:etaQT} simplifies to
\begin{align}
	\eta_q^h(q, \mathcal{T}) &= \hspace{-1em}\sum_{e\,\in\,E_{\mathit{in}}(\mathcal{T})}\mathrm{IsVisible}(e, q), \nonumber \\
\mathrm{IsVisible}(e, q) &= \begin{cases}
	1 & \text{if $(\exists p \in e)[\,\closure{qp} \subset \mathcal{W}\,]$, } \\
	0 & \text{else.}
\end{cases}
\end{align}
The new form of~\eqref{eq:etaT2} is the following:
\begin{equation}
	\eta_\mathcal{T}^h(\mathcal{T}) =
	\sum_{e\,\in\,E_{\mathit{in}}(\mathcal{T})}
	 \iint_{\mathcal{W}}\mathrm{IsVisible}\big(e, (x, y)\big)\cdot f_{X,Y}(x, y)\,\mathrm{d}x\,\mathrm{d}y.
\end{equation}
Applying uniform PDF based on Assumption~\ref{assumption1} results in
\begin{equation}
	\eta_\mathcal{T}^h(\mathcal{T}) =
	 \hspace{-1em}\sum_{e\,\in\,E_{\mathit{in}}(\mathcal{T})}\hspace{-0.5em}\frac{\mathrm{Area}\big(\mathcal{V}_e(e)\big)}{\mathrm{Area}(\mathcal{W})},
	 \label{eq:etaTh}
\end{equation}
where $\mathcal{V}_e$ is the region visible from a segment (see Fig.~\ref{fig:Ve} for the examples):
\begin{equation}
	\mathcal{V}_e(e) = \big\{ v \in \mathbb{R}^2 \;\big\vert\; (\Exists p \in e)[\,\closure{vp} \subset \mathcal{W}\,] \big\}.
	\label{eq:Vs}
\end{equation}
Substituting~\eqref{eq:etaTh} to~\eqref{eq:Tstar} and removing constant factor $1/\mathrm{Area}(\mathcal{W})$, we get
\begin{equation}
	\mathcal{T}^{h\star} = \argmin_{\mathcal{T} \in \mathcal{T}_{\mathcal{W}}(\mathcal{W})}
	\bigg[\;\sum_{e\,\in\,E_{\mathit{in}}(\mathcal{T})}\hspace{-1em}\mathrm{Area}\big(\mathcal{V}_e(e)\big)\;\bigg].
	\label{eq:Tstarfinal}
\end{equation}
We call the resulting mesh $\mathcal{T}^{h\star}$ \emph{minimum visibility triangulation} or MinVT. 





\section{Solution Approach}\label{sec:solution}

\subsection{Minimum Weight Triangulation}
\label{sec:MWT}

\begin{algorithm}[b]
	\caption{MWT for polygon with holes $\mathcal{W}$ with weights $w$.}\label{alg:mwt}
	\begin{algorithmic}[1]
		\Ensure Weight $w_{i,j}$ is precomputed for each pair $(v_i, v_j)$ of $\mathcal{W}$'s vertices. If~$v_i$, $v_j$ are not visible to each other, then it must be that $w_{i,j} = \infty$. 
		\Procedure{MinWeightTriangulation}{$\mathcal{W}$, $w$, $n_\mathcal{P}$, $\mathit{it}_\mathit{max}$, $t_\mathit{max}$}
		\State $\mathcal{T} \leftarrow \Call{ConstrainedDelaunayTriangulation}{\mathcal{W}}$
		\For{$\mathit{it} \leftarrow 1,\dots,\mathit{it}_\mathit{max}$}
		\State\label{algl:mwt:1} $\mathcal{P} \leftarrow \varnothing$ \Comment{simple polygon (to be constructed)}
		\State $\mathcal{T}' \leftarrow \mathcal{T}$ \Comment{mesh with unadded triangles}
		\State $\Lambda \leftarrow \{\text{random triangle${}\in\mathcal{T}'$}\}$ \Comment{set of candidate triangles}
		\While{$\Lambda \neq \varnothing$ and $\vert\mathcal{P}\vert\leq n_\mathcal{P}$}\Comment{$\vert\mathcal{P}\vert\sim{}$the num. of $\mathcal{P}$'s vertices}
		\State $\Delta \leftarrow{}$random triangle${}\in\Lambda$.
		\State $\mathcal{P} \leftarrow \mathcal{P} \cup \Delta$
		\State $\mathcal{T}' \leftarrow \mathcal{T}' \setminus \Delta$
		\State\label{algl:mwt:1.9} $\Lambda \leftarrow \{\Delta' \in \mathcal{T}' \;\vert\;\text{$\mathcal{P}$ and $\Delta'$ share a single edge exclusively}\}$ 
		\EndWhile\label{algl:mwt:2}
		\State\label{algl:mwt:3} $w_\mathcal{P} \leftarrow{}$Reindex weights $w$ according to $\mathcal{P}$'s vertices.
		\State\label{algl:mwt:4} $\mathcal{T}_{\mathcal{P}}\leftarrow{}$\Call{MinWeightTriangulationSimple}{$\mathcal{P}$, $w_\mathcal{P}$}
		\State\label{algl:mwt:5} $\mathcal{T} \leftarrow \mathcal{T}' \cup \mathcal{T}_{\mathcal{P}}$ \Comment{merge the new and original meshes}
		\If{it elapsed more than $t_\mathit{max}$ sec from the procedure's start}
		\State {\bf break}
		\EndIf
		\EndFor
		\State Report $\mathcal{T}$.
		\EndProcedure
	\end{algorithmic}
\end{algorithm}

\begin{figure}[b]
	\centering
	\hfill
	\begin{subfigure}[b]{.19\linewidth}
		\includegraphics[width=\textwidth]{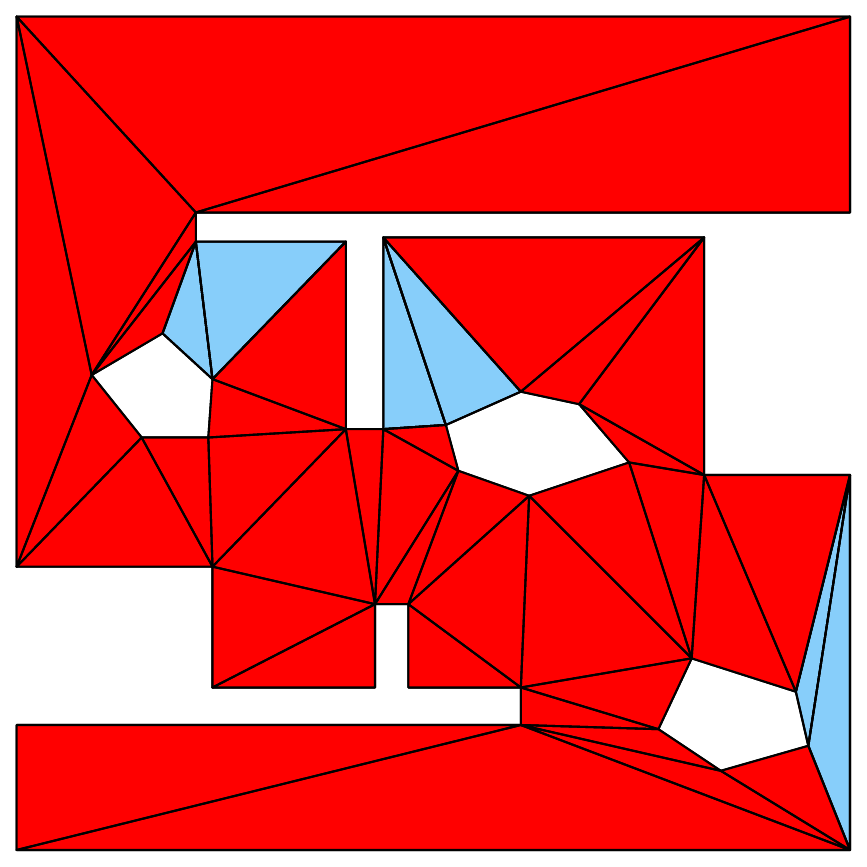}
		\caption*{$\mathcal{T}_{\mathit{prev}},\mathcal{P}$}
	\end{subfigure}
	\hfill
	\begin{subfigure}[b]{.19\linewidth}
		\includegraphics[width=\textwidth]{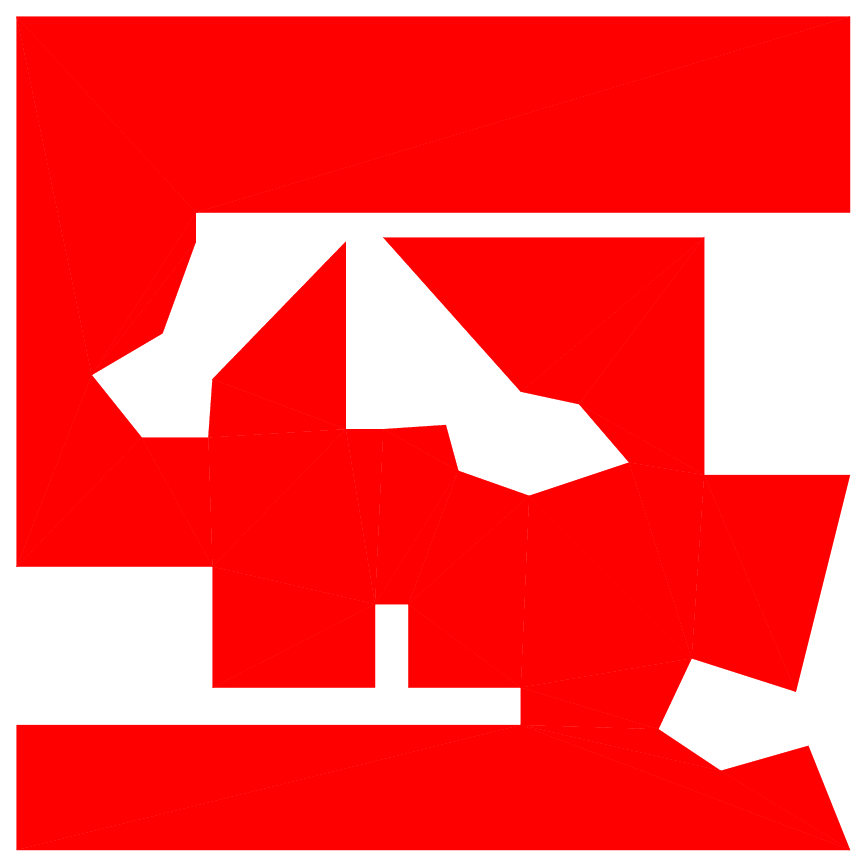}
		\caption*{$\mathcal{P}$}
	\end{subfigure}
	\hfill
	\begin{subfigure}[b]{.19\linewidth}
		\includegraphics[width=\textwidth]{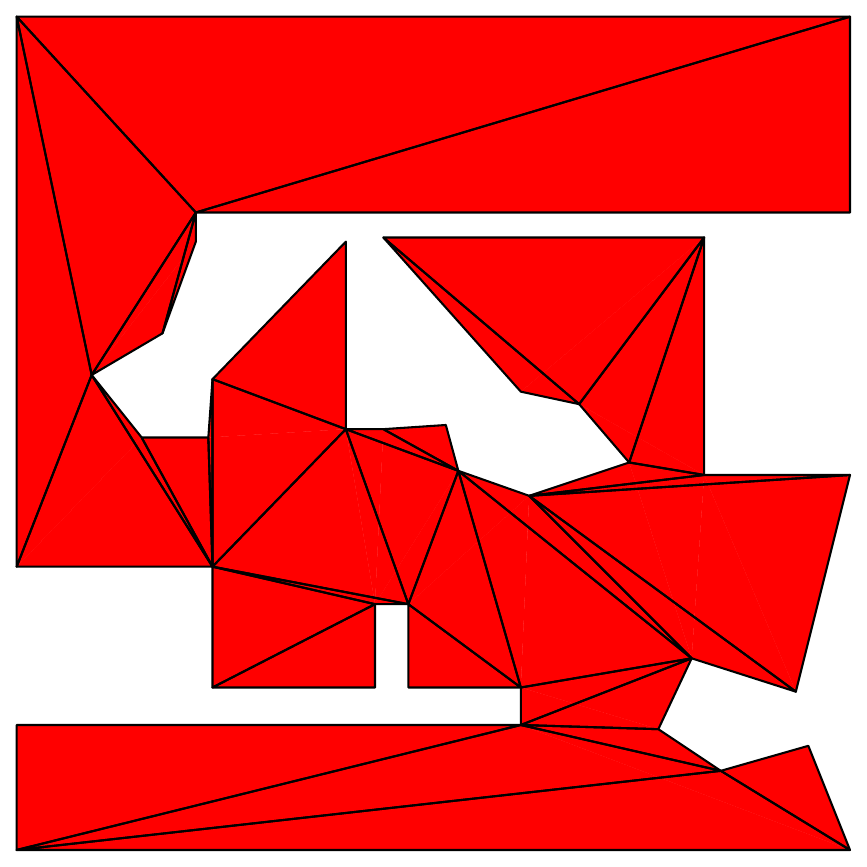}
		\caption*{$\text{MWT}(\mathcal{P})$}
	\end{subfigure}
	\hfill
	\begin{subfigure}[b]{.19\linewidth}
		\includegraphics[width=\textwidth]{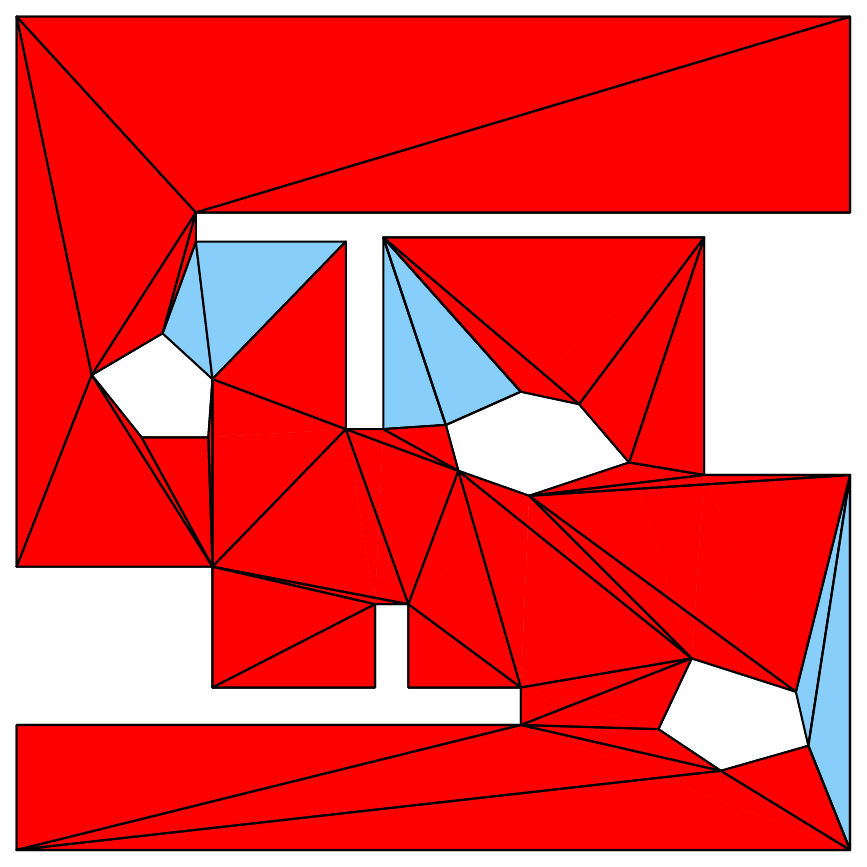}
		\caption*{$\text{MWT}(\mathcal{P}),\mathcal{T}$}
	\end{subfigure}
	\hfill
	\begin{subfigure}[b]{.19\linewidth}
		\includegraphics[width=\textwidth]{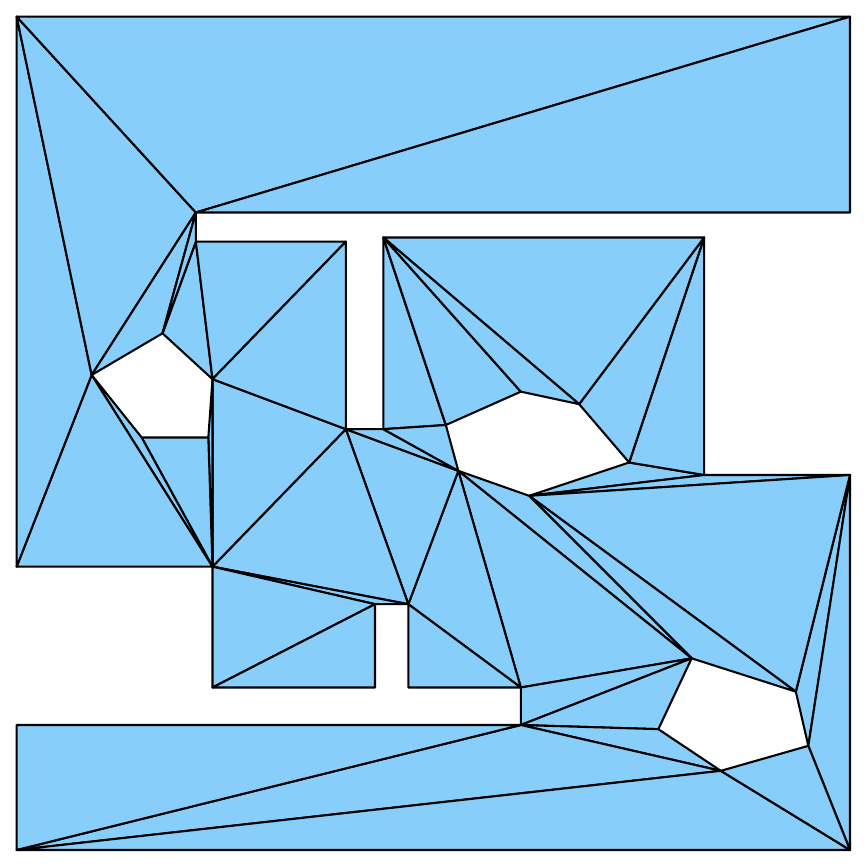}
		\caption*{$\mathcal{T}$}
	\end{subfigure}
	\caption{An example of one iteration of the MWT for a polygon with holes.}
	\label{fig:mwt}
\end{figure}

\begin{algorithm}[b]
	\caption{MWT for simple polygon $\mathcal{P}$ with weights $w$. Based on~\cite{Grantson2008}.}\label{alg:mwt-simple}
	\begin{algorithmic}[1]
		\Ensure $\mathcal{P}$ is simple and its vertices are denoted as $p_1,\dots,p_n$ consecutively around its boundary. The weight of $(p_i,p_j)$ is accessed via $w_{i,j}$. If~$p_i$, $p_j$ are not visible to each other, then it must be that $w_{i,j} = \infty$. 
		\Procedure{MinWeightTriangulationSimple}{$\mathcal{P}$, $w$}
		\For{$i \leftarrow 2,\dots,n$}
		\State $\nu_{i-1,i}\leftarrow 0$
		\State $\kappa_{i-1,i}\leftarrow 0$
		\EndFor
		\For{$\mathit{gap} \leftarrow 2,\dots,n-1$}\label{algl:mwt-simple:5}
		\For{$i \leftarrow 1,\dots,n-\mathit{gap}$}
		\State $j \leftarrow i + \mathit{gap}$
		\State $\nu_{i,j} \leftarrow \infty$
		\For{$k \leftarrow i+1,\dots,j-1$}\label{algl:mwt-simple:9}
		\State\label{algl:mwt-simple:10} $\nu_{\mathit{new}} \leftarrow \nu_{i,k} + \frac{1}{2}(w_{i,j} + w_{j,k} + w_{k,i}) + \nu_{k,j}$
		\If{$\nu_{\mathit{new}} < \nu_{i,j}$}
		\State $\nu_{i,j} \leftarrow \nu_{\mathit{new}}$
		\State\label{algl:mwt-simple:13} $\kappa_{i,j}\leftarrow k$
		\EndIf
		\EndFor\label{algl:mwt-simple:15}
		\EndFor
		\EndFor\label{algl:mwt-simple:17}
		\State\label{algl:mwt-simple:18} \Call{Reconstruct}{$1$, $n$, $\mathcal{P}$}
		\EndProcedure
		\Procedure{Reconstruct}{$i$, $j$, $\kappa$, $\mathcal{P}$}
		\If{$\kappa_{i,j} \neq 0$}
		\State\label{algl:mwt-simple:21} $k \leftarrow \kappa_{i,j}$
		\State\label{algl:mwt-simple:22} Report triangle $\Delta_{i,j,k}$.
		\State\label{algl:mwt-simple:23} \Call{Reconstruct}{$i$, $k$, $\kappa$, $\mathcal{P}$}
		\State\label{algl:mwt-simple:24} \Call{Reconstruct}{$j$, $k$, $\kappa$, $\mathcal{P}$}
		\EndIf
		\EndProcedure
	\end{algorithmic}
\end{algorithm}

The optimization task in Eq.~\eqref{eq:Tstarfinal} is essentially about finding the \emph{minimum weight triangulation} (MWT) for polygon with holes $\mathcal{W}$, where the weights are somehow specifically defined. 
Finding the MWT for a general set of points on a plane and arbitrarily defined weights is an NP-hard problem~\cite{Mulzer2008}. 
The MWT for a polygon with holes is the constrained version of the MWT for points, similarly as the CDT is the \emph{constrained} version of the \emph{Delaunay triangulation}.
The MWT for polygon with holes $\mathcal{W}$ always contains edges that are part of the $\mathcal{W}$'s boundary. 
Also, we are interested only in triangulating $\mathcal{W}$'s interior, not the outside.
For computing the MWT for simple polygons, there are known algorithms that are straightforward to implement~\cite{Grantson2008}. 
However, we found no such convenient algorithm for polygons with holes. 
Therefore we propose our own heuristic method in~Alg.~\ref{alg:mwt}.

The method is based on an iterative improvement of the $\mathcal{W}$'s CDT with respect to provided weights $w$. 
In each iteration, simple polygon $\mathcal{P}$ is generated in a randomized fashion by adding triangles from the mesh from the previous iteration (lines~\ref{algl:mwt:1}-\ref{algl:mwt:2}). 
The simplicity of $\mathcal{P}$ is assured by adding only triangles that share a single edge with $\mathcal{P}$ exclusively (line~\ref{algl:mwt:1.9}). 
If a triangle shares no edge, more than one edge, or one edge and a vertex other than the edge's endpoints, then the triangle is not considered for addition. 
This way, no hole can be ever added to~$\mathcal{P}$. 
The iteration ends by computing the MWT for~$\mathcal{P}$ with a method from~\cite{Grantson2008} and replacing the triangles added to~$\mathcal{P}$ in the previous iteration mesh with the optimal triangles from MWT, forming the new mesh (lines~\ref{algl:mwt:3}-\ref{algl:mwt:5}). 
An example of one iteration is shown in Fig.~\ref{fig:mwt}.

The MWT for simple polygon~$\mathcal{P}$ is composed of recursive substructure that allows employing \emph{dynamic programming} from~\cite{Grantson2008}, shown in~Alg.~\ref{alg:mwt-simple}.  
Assuming that $\mathcal{P}$'s vertices are numbered in cw or ccw order around its boundary, the method builds tree of weights $\nu_{i,j}$ associated with subpolygons with vertices between $p_i$ and $p_j$ (lines~\ref{algl:mwt-simple:5}-\ref{algl:mwt-simple:17}). 
The tree is build starting from triangles and is finished when $\nu_{1,n}$ is determined, since indices~$(1,n)$ correspond to the original polygon.
The weights of non-trivial subpolygons defined by indices~$(i,j)$ are determined by evaluating all possible triangles $\Delta_{i,j,k}$ within them (lines~\ref{algl:mwt-simple:9}-\ref{algl:mwt-simple:15}). 
Each triangle divides the subpolygon into three parts: the sub-subpolygon to the left, the triangle itself, and the sub-subpolygon to the right (line~\ref{algl:mwt-simple:10}). 
Weight $\nu_{i,j}$ is determined as the minimal sum of these parts, while index $k$ that resulted in the minimal sum is stored in $\kappa_{i,j}$ (line~\ref{algl:mwt-simple:13}).
In the end, the resulting MWT for $\mathcal{P}$ is reconstructed from the $\kappa$-indices.  
The~tree structure of $\kappa$ is recursively traversed starting from $(1,n)$ (line~\ref{algl:mwt-simple:18}). 
In each step, pair $(i,j)$ determines triangle $\Delta_{i,j,\kappa_{i,j}}$, which is reported as part of the resulting mesh (lines~\ref{algl:mwt-simple:21}-\ref{algl:mwt-simple:22}). 
The reconstruction is then called for $\Delta_{i,j,\kappa_{i,j}}$'s edges $(i, \kappa_{i,j})$ and $(j, \kappa_{i,j})$ until all triangles are eventually reported~(lines~\ref{algl:mwt-simple:23}-\ref{algl:mwt-simple:24}).

The whole framework for constructing MWT for polygon with holes, i.e., Alg.~\ref{alg:mwt}-\ref{alg:mwt-simple}, has one parameter and two stopping conditions. 
The parameter is the maximum number of vertices $n_\mathcal{P}$ of simple polygon $\mathcal{P}$.
If $\mathcal{P}$ reaches this size, its construction is finished prematurely. 
But setting $n_\mathcal{P}\;{=}\;\infty$ is also valid. 
$\mathcal{P}$ is then always maximal such that adding any other unadded triangle would break its simplicity. 
The stopping conditions are the maximal number of iterations $\mathit{it}_\mathit{max}$ and the maximal computational time $t_\mathit{max}$, and they are evaluated in the main loop of Alg.~\ref{alg:mwt}.

\subsection{Minimum Visibility Triangulation}
\label{sec:MinVT}

The difference between the MWT and MinVT is the definition of weights for every pair of $\mathcal{W}$'s vertices $(v_i, v_j)$.
While the MWT assumes the weights are given, the MinVT includes their exact definition.
This subsection primarily concerns how these weights are computed.
But first, the following paragraph summarizes some general information about the weights, including the notation used in the rest of this paper. 

It is equivalent to say that $w_{i,j}$ is defined for the pair of $\mathcal{W}$'s vertices $(v_i, v_j)$ as to say that it is defined for the segment $\closure{v_iv_j}$.
The~segment $\closure{v_iv_j}$ can also be called \emph{edge} $e_{i,j}$. 
For~Alg.~\ref{alg:mwt}-\ref{alg:mwt-simple} to produce a valid triangular mesh, the following is required of the weights: 
\begin{equation}
	w_{i,j} = w_{j,i} = \begin{cases}
		\infty & \text{if }\closure{v_iv_j} \not\subset \mathcal{W}, \\
		0 & \text{if }\closure{v_iv_j} \subset \mathrm{Bd}(\mathcal{W}), \\
		\text{any finite value} & \text{else.}
	\end{cases}
	\label{eq:w-general}
\end{equation}
Note that the weights must be symmetrical.
The first case of~\eqref{eq:w-general} refers to the \emph{forbidden edges}, which can never appear in any valid triangulation because their endpoints are not visible to each other; therefore, they have infinite weights.  
The second case concerns the \emph{boundary edges}, which must be present in every valid mesh of $\mathcal{W}$ and, by convention, they have zero weights.  
The last case is about \emph{interior edges} which may be present in a valid triangulation and sometimes must; therefore, their value must be finite. 
Only the interior edges are the ones that are optimized and their weight value matters to the optimization result. 
\emph{Thereinafter, when we speak about a weight of an edge, we implicitly mean the edge is an interior edge.}

MinVT's weights are defined according to~\eqref{eq:Tstarfinal} as $w_{i,j} = \mathrm{Area}\big(\mathcal{V}_e(e_{i,j}))$, where
$\mathcal{V}_e(e)$ is the region visible from edge $e$ defined in Eq.~\eqref{eq:Vs}. 
This~region is a structure similar to the visibility region from a point, but as if the point traversed along the segment from one endpoint to the other, gathering all points that were visible at some point during that journey.
We use an approximate computation that utilizes the classic TEA running on an un-optimized CDT mesh.
The computation samples the segment $e = \closure{uv}$ in between its endpoints equidistantly with points $S(e, d_{\mathit{samp}}) = \{ s_1, s_2, \dots \}$ such that the sampling distance is maximal but not larger than parameter $d_{\mathit{samp}}$, unless $\vert e \vert \leq d_{\mathit{samp}}$.
If the latter condition holds, then just one sample in the middle of the edge is used instead, i.e., $S(e,d_{\mathit{samp}}) = \{s_1=(u+v)/2\}$.
The desired region is then obtained as the union of the visibility regions from all points in $S$:
\begin{equation}
	\mathcal{V}_e(e) = \hspace{-1em} \bigcup_{s\,\in\,S(e,d_{\mathit{samp}})} \hspace{-1em} \mathcal{V}(s).
	\label{eq:Vssamp}
\end{equation}

\subsection{Limited Visibility Range}
\label{sec:d}

\begin{figure}[b]
	\centering
	\begin{subfigure}[t]{0.19\textwidth}
		\begin{overpic}[width=\columnwidth]{g/visibility_region.pdf}
			\put(3,50){$\mathcal{W}$}
			\put(46,37){$q$}
			\put(33,70){$\mathcal{V}(q)$}
		\end{overpic}
	\end{subfigure}
	\hspace{1em}
	\begin{subfigure}[t]{0.19\textwidth}
		\begin{overpic}[width=\columnwidth]{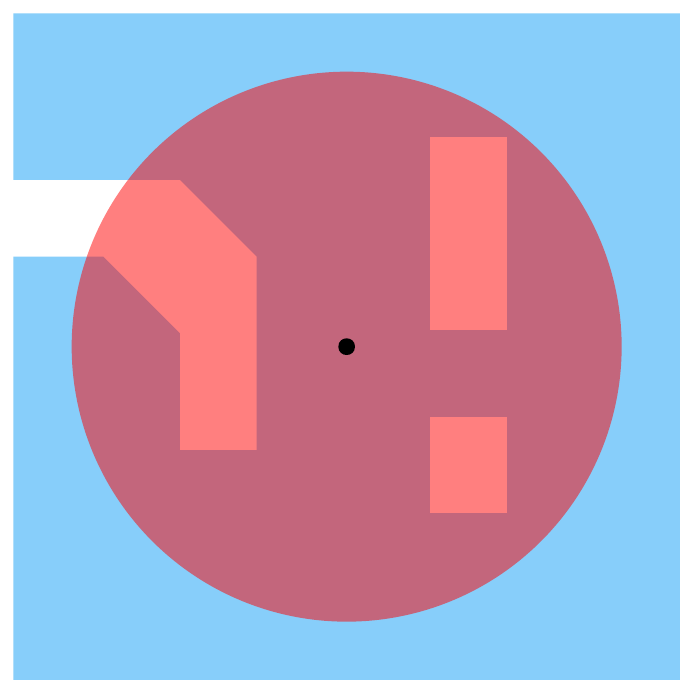}
			\put(3,87){$\mathcal{W}$}
			\put(19,68){$\mathit{Circle}(q, d)$}
			\put(46,37){$q$}
		\end{overpic}
	\end{subfigure}
	\hspace{1em}
	\begin{subfigure}[t]{0.19\textwidth}
		\begin{overpic}[width=\columnwidth]{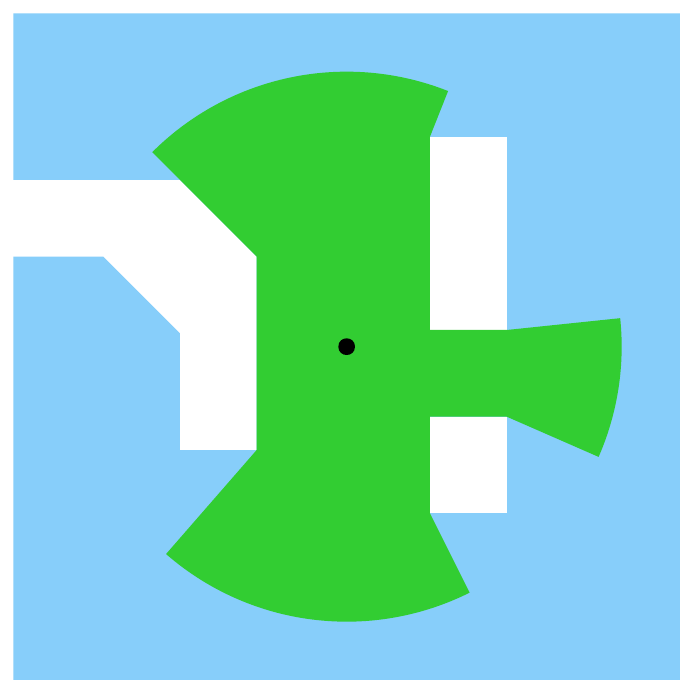}
			\put(3,87){$\mathcal{W}$}
			\put(46,37){$q$}
			\put(23,77){$\mathcal{V}(q, d)$}
		\end{overpic}
	\end{subfigure}
	\caption{An example of a visibility region with limited range $d$.}
	\label{fig:limited-vis}
\end{figure}
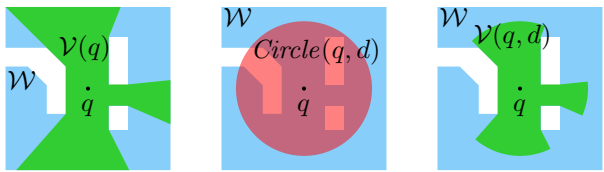

Visibility regions can be thought of as a general concept to model an agent's ability to see or cover part of the environment in which it operates. 
The visibility regions considered so far are the most idealized case of an omnidirectional vision with an unlimited visibility range. 
However, visibility regions can be subject to additional constraints in real-life applications. 
For example, in robotics, omnidirectional sensors are common, but their resolution over longer distances is always limited to finite range~$d$. 
If we wanted to model this situation, we could run the TEA to get the standard visibility region and then intersect it with a circle.
See the example in Fig.~\ref{fig:limited-vis}.
The result can be formally defined as
\begin{equation}
\mathcal{V}(q,d) = \big\{ v \in \mathbb{R}^2 \;\big\vert\; (\forall v)[\, \closure{vq} \subset \mathcal{W} \text{ and } \vert\closure{vq}\vert \leq d \,] \big\}.
\end{equation}
However, if the result was considerably smaller than the original, we would have wasted much computational power when TEA expanded beyond range~$d$. 

To be more efficient, we suggest an early exit strategy for the TEA. 
The~only change is that edges that do not intersect the circle of radius~$d$ centered in the query point are handled the same way as boundary edges. 
We call this modification d-TEA. 
Note that TEA = $\infty$-TEA. 
The~output of d-TEA still needs to undergo the intersection with the circle to get $\mathcal{V}(q,d)$, but the computational advantage of d-TEA is that it expands only edges that contribute to the final result. 
The circle intersection can be computed efficiently by rotating around the query point, taking advantage of the region's star shape. 

\begin{figure}
	\centering
	\begin{subfigure}[b]{.19\linewidth}
		\includegraphics[width=\textwidth]{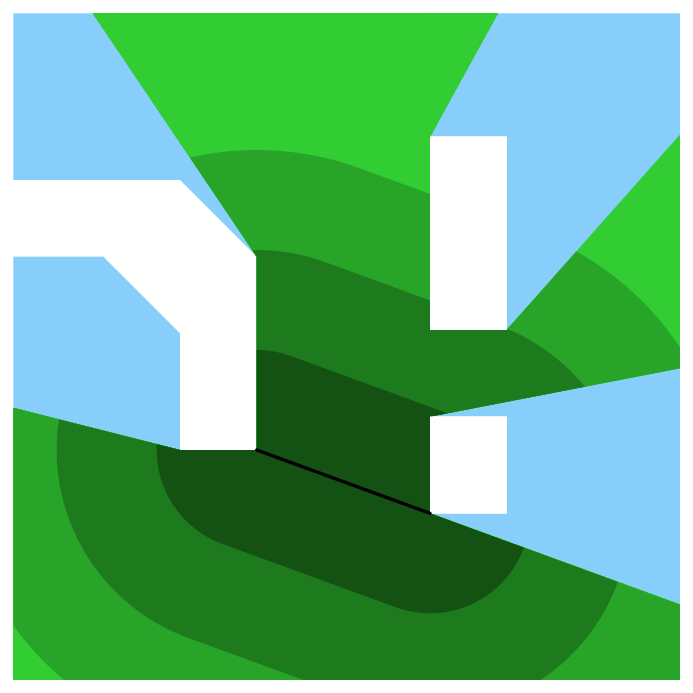}
	\end{subfigure}
	\hfill
	\begin{subfigure}[b]{.19\linewidth}
		\includegraphics[width=\textwidth]{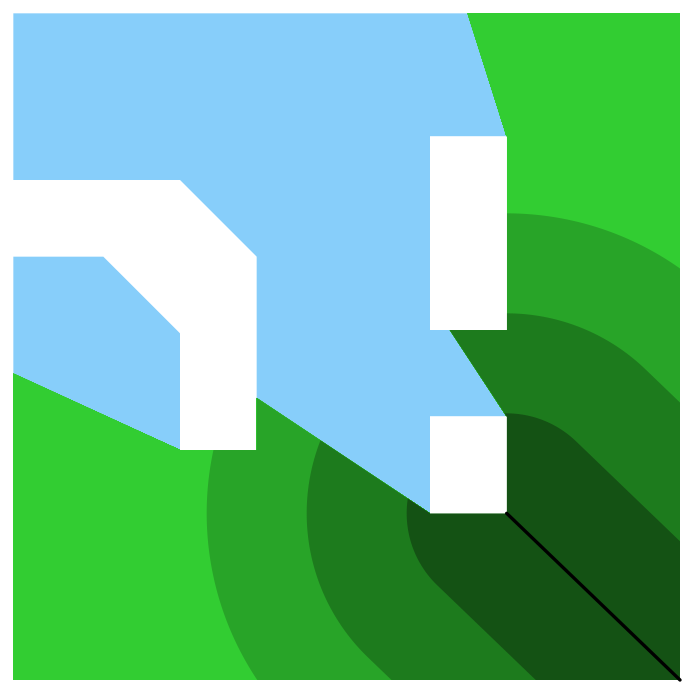}
	\end{subfigure}
	\hfill
	\begin{subfigure}[b]{.19\linewidth}
		\includegraphics[width=\textwidth]{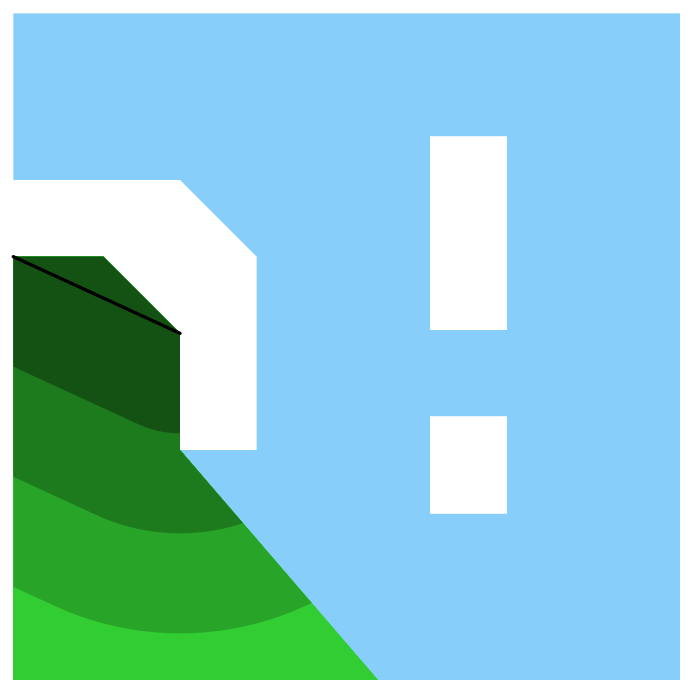}
	\end{subfigure}
	\hfill
	\begin{subfigure}[b]{.19\linewidth}
		\includegraphics[width=\textwidth]{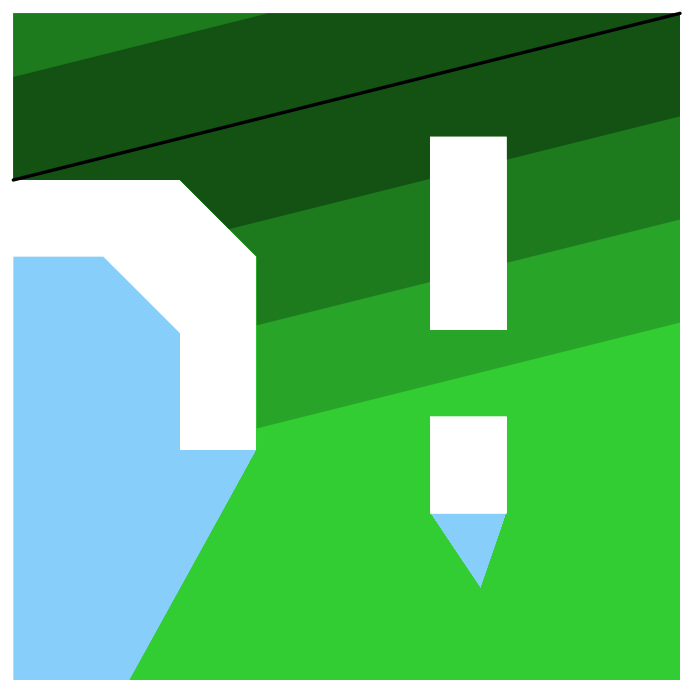}
	\end{subfigure}
	\hfill
	\begin{subfigure}[b]{.19\linewidth}
		\includegraphics[width=\textwidth]{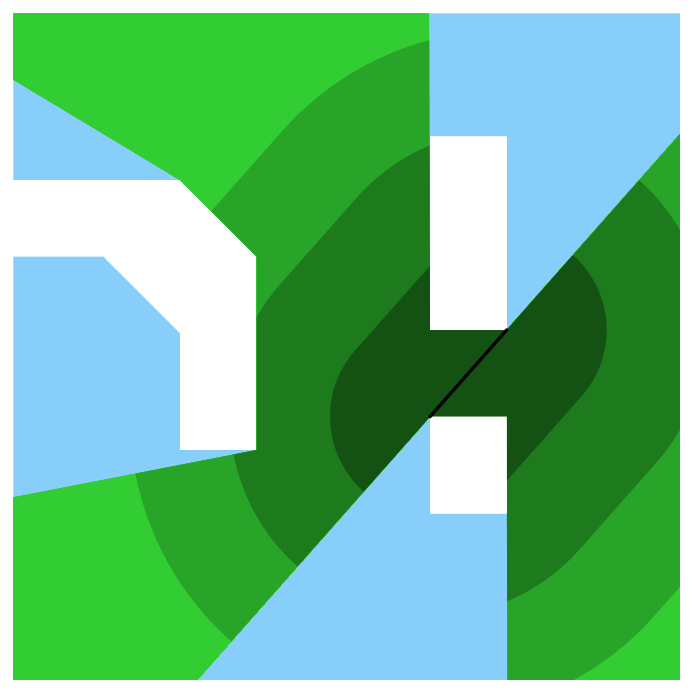}
	\end{subfigure}
	\caption{Examples of regions $\mathcal{V}_e(e,d)$ visible from segments $e$ for different $d$s.}
	\label{fig:Ved}
\end{figure}

While it may appear trivial, extending from TEA to d-TEA alters the number of expanded edges and consequently influences the optimal mesh, contingent on the value of $d$. 
To facilitate this adjustment, we integrated $d$ into each equation from \eqref{eq:etaT} to \eqref{eq:Tstarfinal}. Nevertheless, for the sake of brevity, we provide only the final result here.
We denote the d-TEA's equivalent to MinVT as d-MinVT, and define its weights as $w_{i,j} = \mathrm{Area}\big(\mathcal{V}_e(e_{i,j},d))$, where
\begin{equation}
	\mathcal{V}_e(e,d) = \big\{ v \in \mathbb{R}^2 \;\big\vert\; (\Exists p \in e)[\,\closure{vp} \subset \mathcal{W} \text{ and } \vert\closure{vp}\vert \leq d\,] \big\}.
	\label{eq:Vsd}
\end{equation}
See Fig.~\ref{fig:Ved} for the examples.
To evaluate d-MinVT, we compute $\mathcal{V}_e(e,d)$ through the following process. 
Initially, we compute $\mathcal{V}'_e(e)$ using d-TEA instead of TEA, following the formula in~\eqref{eq:Vssamp}. 
Subsequently, we intersect $\mathcal{V}'_e(e)$ with a shape formed by the union of the following components: two circles with a radius of $d$, each centered at one of $e$'s endpoints, and a belt of width $2d$ extending from one endpoint to the other.

\subsection{T\v{r}iVis: The Implementation}
\label{sec:imp}

Our \texttt{C++} implementation of the TEA, d-TEA, MWT, and MinVT, is publicly available\footnote{\url{http://imr.ciirc.cvut.cz/Research/TriVis}}, along with the data and scripts to run our experiments.
We call the implementation T\v{r}iVis\footnote{
    In the initial conference paper~\cite{Mikula2022b}, we referred to the implementation as TriVis. 
    However, we have since renamed it to T\v{r}iVis, signifying the third version, where \emph{t\v{r}i} translates to \emph{three} in Czech. 
    Letter \v{r} is pronounced as 'rzh' with a rolled 'r'.
}.
T\v{r}iVis is a robust, versatile, easy-to-use, self-contained, open-source \texttt{C++} library for computing visibility-related structures. 

The robustness is ensured using Shewchuck's adaptive robust orientation predicates~\cite{Shewchuk1997} in the TEA expansion procedure.  
In addition to the TEA for computing visibility regions, it includes TEA variants optimized for determining visibility between two query points and for computing all vertices of the map visible from a given point, all with or without considering the limited visibility range.

The visibility regions come with edge and vertex flags, indicating their origin. 
For example, an edge of a visibility region can either be part of the map boundary or free space. 
Similarly, a vertex can either be a vertex of the map or computed as a result of a ray-segment intersection. 
This information can be useful when the topological relationship between the region and the map is crucial for users.

Thanks to the initial abstract representation of the visibility region, the TEA expansion procedure and the computation of ray-segment intersections are separated in the implementation. 
Therefore, users can save runtime by skipping the computation of intersections if the abstract representation is sufficient for their needs.

Locating the triangle that contains the query point is implemented efficiently. 
It~can be done by a simple walk in $\mathcal{O}(n)$, as Bungiu~et~al. \cite{Bungiu2014} suggest, but we do it with a more sophisticated technique called \emph{bucketing}~\cite{Fiser2013,Edahiro1984}, which uses rectangular regions called \emph{buckets}.
The bucket containing $q$ is identified first by simply rounding the coordinates of $q$ and using a lookup table.
The~bucket points to all triangles in $\mathcal{T}$ that share a non-zero intersection with it, so $\Delta_q$ is found by checking only these triangles. 
Bucketing requires another round of preprocessing to determine the intersecting triangles for each bucket and generally does not improve the worst-case time complexity, which remains $\mathcal{O}(n)$ to find $\Delta_q$.\footnote{
	This is because a bucket can still intersect $\mathcal{O}(n)$ triangles in the general case. 	
}
However, in practice, the expected run time is $\mathcal{O}(1)$ for most meshes when the bucket size is reasonably chosen.

\section{Experiments, Results, and Discussion}\label{sec:results}

This section describes and discusses all the experiments and results to evaluate our assumptions (\ref{sec:eval-assumptions}), tune our algorithms (\ref{sec:tuning}), and thoroughly test our approach (\ref{sec:eval-tea}-\ref{sec:eval-dtea}). 
All these experiments have two things in common: the type of benchmark instances and hardware settings. 
The benchmark instances are the complex polygonal maps from~\cite{Harabor2022} based on the video game Iron Harvest from KING Art Games. 
From the set of 35, we randomly select 10 of them for tuning purposes and use the rest for the final evaluation. 
Appx.~\ref{sec:maps} provides more information about the maps. 
All the experiments are executed on a personal laptop Legion~5~Pro 16ITH6H with Intel Core i7-11800H (4.60~GHz), 16~GB of RAM, and running Ubuntu 20.04.5 LTS. 
The implementation is single-threaded.

Additionally, through this section, we often refer to the \emph{relative value} or \emph{percentage gap} of some variable.
Both refer to the same definition:
\begin{equation}
	\%\mathit{val} = 100 \cdot (\mathit{val} - \mathit{val}_\mathit{ref})\;/\;\mathit{val}_\mathit{ref},
	\label{eq:gap}
\end{equation}
where $\mathit{val}$ is the absolute value and $\mathit{val}_\mathit{ref}$ is the reference absolute value of the same type.
In all cases, we use the CDT as the reference.

\subsection{Evaluation of the Assumptions}
\label{sec:eval-assumptions}

\begin{figure}[b]
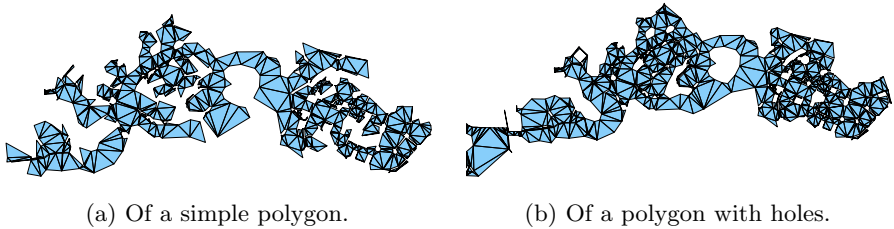

	\centering
	\begin{subfigure}[b]{0.49\textwidth}
		\centering
		\includegraphics[width=1.0\textwidth]{g/assumption_map_simple}
		\caption{Of a simple polygon.}
		\label{fig:mesh-simple}
	\end{subfigure}
	\hfill
	\begin{subfigure}[b]{0.49\textwidth}
		\centering
		\includegraphics[width=1.0\textwidth]{g/assumption_map_with_holes}
		\caption{Of a polygon with holes.}
		\label{fig:mesh-holes}
	\end{subfigure}
	\caption{An example of triangular mesh.}
	\label{fig:meshes-simple-holes}
\end{figure}

This section experimentally shows that the assumptions from Sec.~\ref{sec:problem} approximately hold.
First, we show that the computational time $t_q$ is approximately proportional to the number of expansions $\eta_q$ for our experimental setup. 
Second, we address the relationship between $\eta_\mathcal{T}$ and $\eta_\mathcal{T}^h$ from Eq.~\eqref{eq:etaT2} and~\eqref{eq:etaTh}, respectively, for a simple polygon and polygon with holes. 
In both scenarios we use the two mesh instances shown in Fig.~\ref{fig:meshes-simple-holes}.

\subsubsection{Relation Between $\eta_q$ and $t_q$}
\label{sec:exp-ass-time}

\begin{figure}
	\begin{subfigure}[b]{0.49\textwidth}
		\centering
		\includegraphics[width=1.0\textwidth]{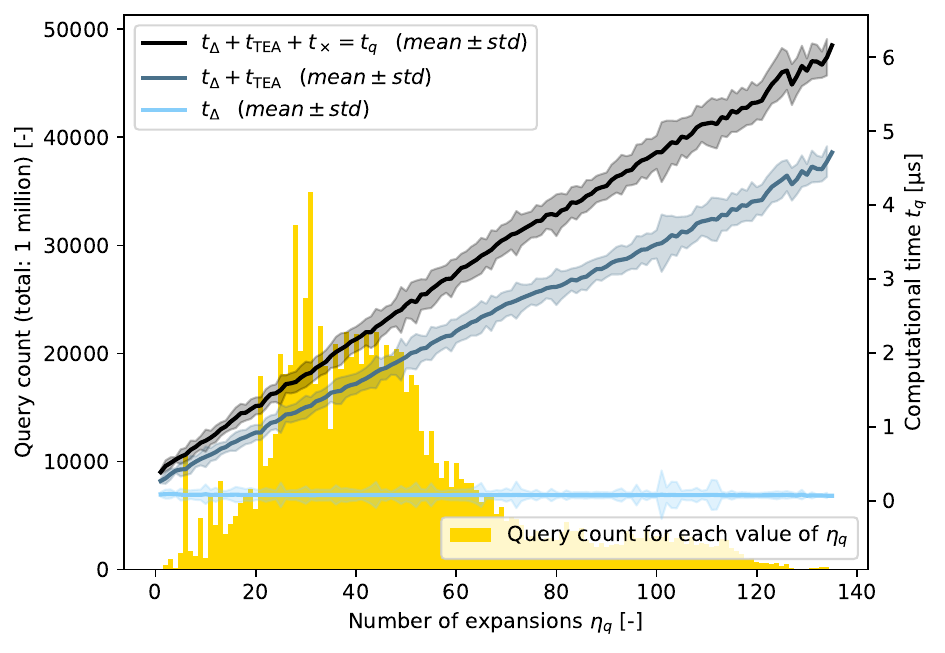}
		\caption{For the mesh from Fig.~\ref{fig:mesh-simple}.}
	\end{subfigure}
	\hfill
	\begin{subfigure}[b]{0.49\textwidth}
		\centering
		\includegraphics[width=1.0\textwidth]{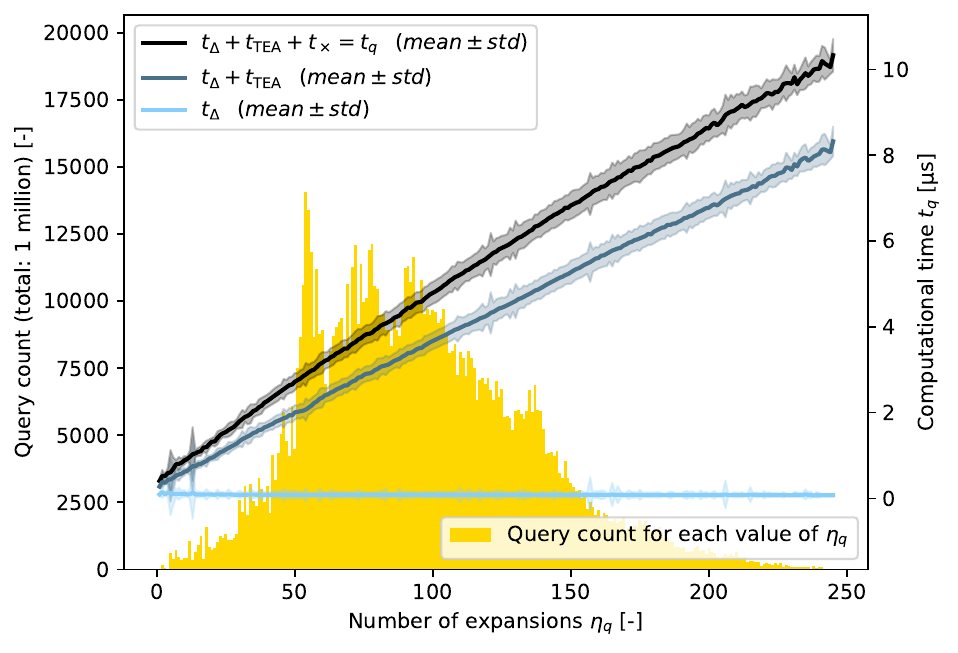}
		\caption{For the mesh from Fig.~\ref{fig:mesh-holes}.}
	\end{subfigure}
	\caption{Experimental evaluation of the relation between $\eta_q$ and $t_q$.}
	\label{fig:expansions-time}
\end{figure}

We have run the TEA on two meshes to find visibility regions for one million uniformly random query points, measuring both metrics. 
For each value of~$\eta_q$, Fig.~\ref{fig:expansions-time} shows how many times this value was recorded (gold) and the corresponding mean value $\pm$ standard deviation of $t_q$ (black).
Our implementation allows to measure $t_q$ as the sum of three consecutive times: $t_q = t_\Delta + t_\text{TEA} + t_\times$, where $t_\Delta$ is the time to locate $\Delta_q$, $t_\text{TEA}$ is the time of the expansion procedure without computing the ray-segment intersection points between edges and $q$'s views, and $t_\times$ is the time to compute all these intersection points. 
The figures furthermore show the plots of $t_\Delta$ (light blue) and $t_\Delta + t_\text{TEA}$ (dark blue). 
We can see that the expansion procedure with time $t_\text{TEA}$ is the most computationally demanding part. 
Time $t_\Delta$ is constant and very small, while both $t_\text{TEA}$ and $t_\times$ depend almost linearly on the number of expansions. 
This observation is expected for~$t_\text{TEA}$, but it is not obvious for $t_\times$, because this part logically depends purely on the output and not on the mesh.
A possible explanation is that more expansions usually relate to a more complex output; thus, more intersection points must be computed. 

The overall conclusion is that the total time $t_q$ to compute visibility region is approximately proportional to the number of expansions $\eta_q$ in the case of our experimental setup. 
Minimizing~$\eta_q$ to improve $t_q$ is thus a solid idea.


\subsubsection{Relation Between $\eta_\mathcal{T}$ and $\eta_\mathcal{T}^h$}  
\label{sec:exp-ass-weights}

\begin{figure}
	\begin{subfigure}[b]{0.49\textwidth}
		\centering
		\includegraphics[width=1.0\textwidth]{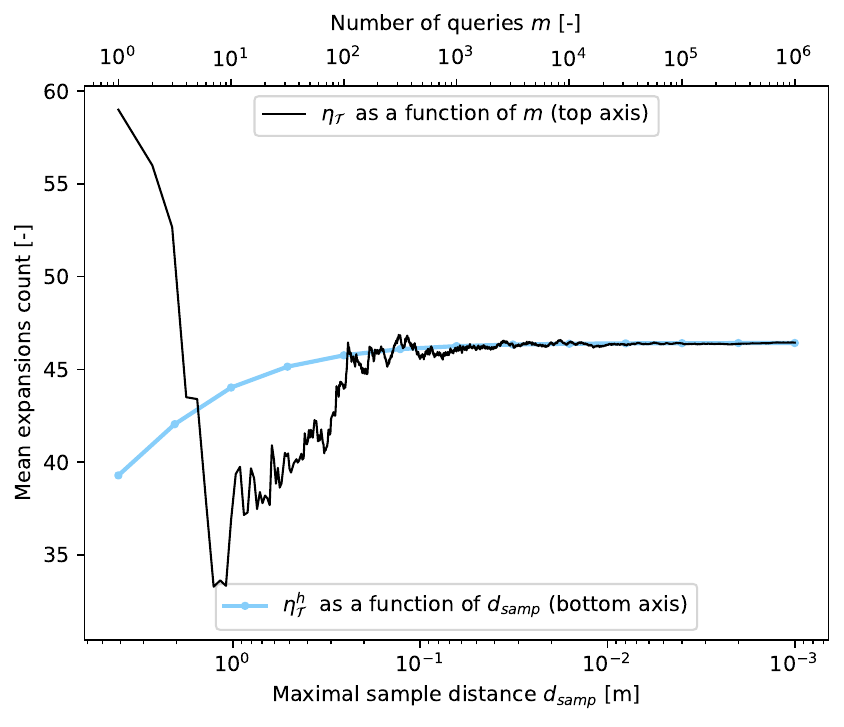}
		\caption{For the mesh from Fig.~\ref{fig:mesh-simple}.}
		\label{fig:assumption-weights-a}
	\end{subfigure}
	\hfill
	\begin{subfigure}[b]{0.49\textwidth}
		\centering
		\includegraphics[width=1.0\textwidth]{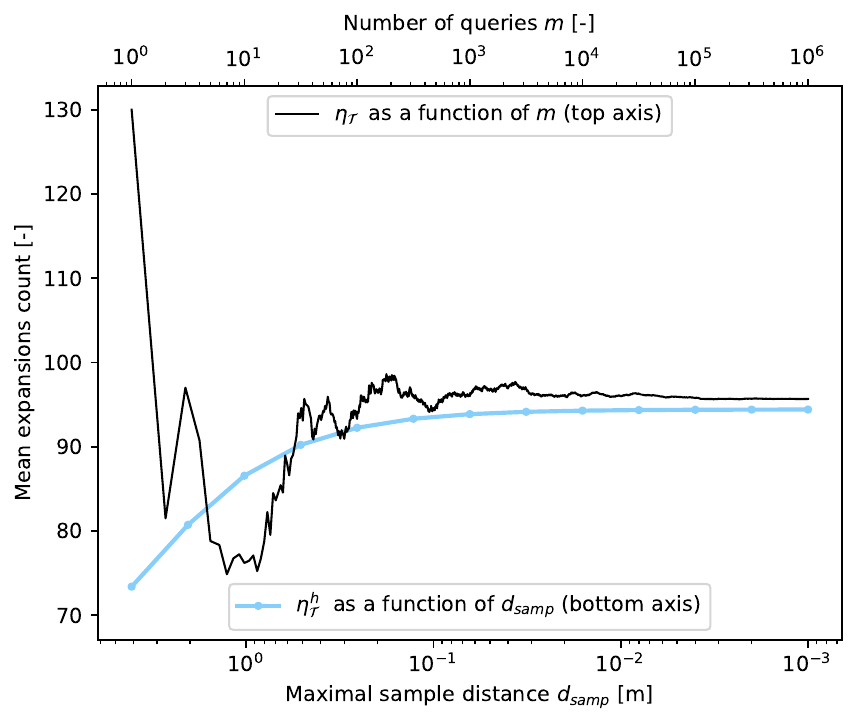}
		\caption{For the mesh from Fig.~\ref{fig:mesh-holes}.}
		\label{fig:assumption-weights-b}
	\end{subfigure}
	\caption{Experimental evaluation of the relation between $\eta_\mathcal{T}$ and $\eta_\mathcal{T}^h$.}
	\label{fig:assumption-weights}
\end{figure}

While $\eta_\mathcal{T}$ is the expected number of edge expansions, $\eta_\mathcal{T}^h$ is the expected number of expanded edges. 
The optimal mesh for the TEA wishes to minimize the first, but our heuristic approach minimizes the second with MinVT. 
The~process of deriving $\eta_\mathcal{T}^h$ (Sec.~\ref{sec:problem-assumptions}) implies that if the assumption~\eqref{eq:assumption} holds for assessed~$\mathcal{W}$, then $\eta_\mathcal{T} = \eta_\mathcal{T}^h$.
We know that \eqref{eq:assumption} always holds for simple polygons, but it may not hold for polygons with holes. 
Since none of the practical instances we solve later in the evaluation are simple, we must hope that for them \eqref{eq:assumption} holds approximately. 
The following experiment addresses these issues.

The TEA was run for $m$ uniformly random query points while recording the numbers of expansions, which were then averaged to get the estimate of~$\eta_\mathcal{T}$.
The estimate of $\eta_\mathcal{T}^h$ was obtained according to Eq.~\eqref{eq:etaTh} and \eqref{eq:Vssamp} for several values of sampling distance $d_{\mathit{samp}}$.
Fig.~\ref{fig:assumption-weights} shows~$\eta_\mathcal{T}$ (black) as the function of $m$ (top axis), and $\eta_\mathcal{T}^h$ (light blue) as the function of $d_{\mathit{samp}}$ (bottom axis).
We~can see that the two plots converge to the same value in the case of the simple polygon (Fig.~\ref{fig:assumption-weights-a}), according to our expectations.\footnote{
    The initial variation in $\eta_\mathcal{T}$ results from the randomness in sample selection, whereas for $\eta_\mathcal{T}^h$, it arises due to the initial poor quality of estimation (attributed to too high values of $d_{\mathit{samp}}$).
}
The~final gap of $\eta_\mathcal{T}^h$ from $\eta_\mathcal{T}$ is only $-0.04\%$. 
In the case of the polygon with holes, the two plots also converge to a value, but each to a slightly different one (Fig.~\ref{fig:assumption-weights-b}).
The~final gap (of $-1.3\%$) was recorded because some edges were expanded more than once in the evaluation, which is not coped by the value of $\eta_\mathcal{T}^h$. 
The gap can be seen as the estimate of how imprecise our approach is when computing the optimization criterion inside a polygon with holes representing a realistic yet complex environment. 

\subsection{Construction Tuning}
\label{sec:tuning}

This subsection aims to find the best balance between the quality of the MinVT mesh and the computational time required for its construction. 
The objective is to construct a mesh of sufficient quality without unnecessarily consuming computational resources. 
Table~\ref{tab:params} presents all tunable parameters of the construction procedure that influence both quality and construction time.
It is observed that certain parameter values that enhance mesh quality also result in a slowdown in construction time. 
The subsequent sections detail the experiments that guided us in selecting specific parameter values, shown in the 'Value' column, for the final evaluation.
It's important to note that the map instances used for tuning the parameters are distinct from those employed in the final evaluation. 
In this context, we utilize only the 10 instances marked with \cmark{} in Table~\ref{tab:maps}. 
The subsequent evaluation sections utilize the remaining 25 unmarked instances.

\begin{table}[t]
	\begin{center}
		\begin{minipage}{\textwidth}
			\caption{Tunable parameters of the MinVT construction.}\label{tab:params}%
			\begin{tabular*}{\textwidth}{@{\extracolsep{\fill}}lllcc@{\extracolsep{\fill}}}
				\toprule
				Symbol & Where introduced & Meaning & Value & Unit \\
				\midrule
				$d_{\mathit{samp}}$    & Sec.~\ref{sec:MinVT},~Eq.~\eqref{eq:Vssamp}   & Maximal sampling distance for $\mathcal{V}_e$ & $\{2,4,8\}$ & m \\
				$n_\mathcal{P}$    & Sec.~\ref{sec:MWT},~Alg.~\ref{alg:mwt}   & Maximal size of simple polygon $\mathcal{P}$  & 450 & 1 \\
				$\mathit{it}_\mathit{max}$ & Sec.~\ref{sec:MWT},~Alg.~\ref{alg:mwt}   & Maximal number of MWT iterations  & 200 & 1 \\
				$t_\mathit{max}$    & Sec.~\ref{sec:MWT},~Alg.~\ref{alg:mwt}   & Maximal run-time of MWT & 6 & s \\
				$r_\mathit{pen}$    & The end of Sec.~\ref{sec:tuning}  & How many edges are penalized  & $\{0,50\}$ & \% \\
				\botrule
			\end{tabular*}
		\end{minipage}
	\end{center}
\end{table}

\begin{figure}[b]
	\centering
	\begin{subfigure}[b]{0.49\textwidth}
		\centering
		\includegraphics[width=1.0\textwidth]{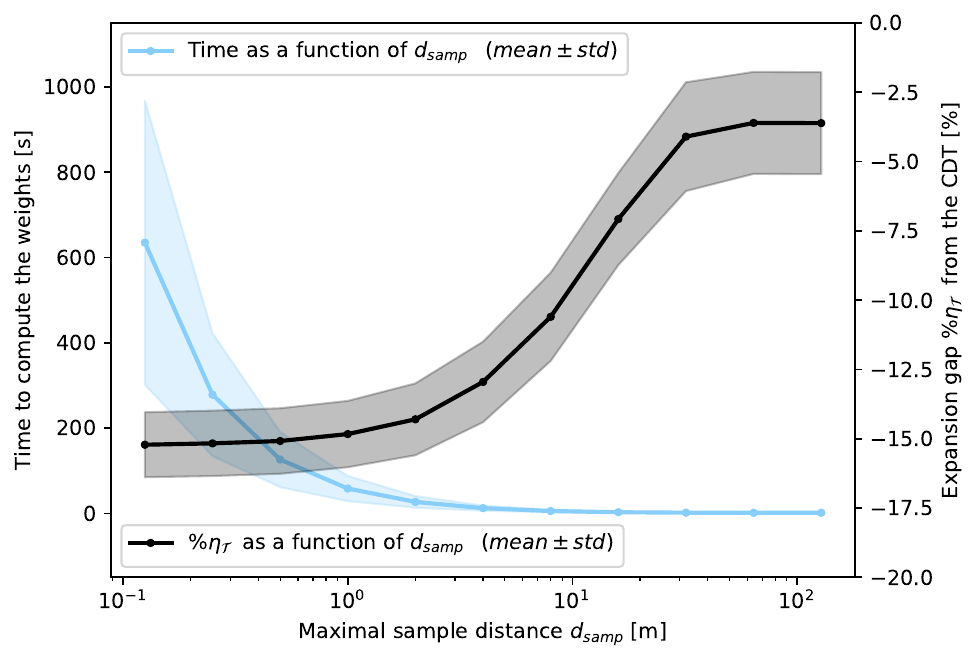}
		\caption{Tuning the parameter $d_{\mathit{samp}}$.}
		\label{fig:eval-sample-dist}
	\end{subfigure}
	\hfill
	\begin{subfigure}[b]{0.49\textwidth}
		\centering
		\includegraphics[width=1.0\textwidth]{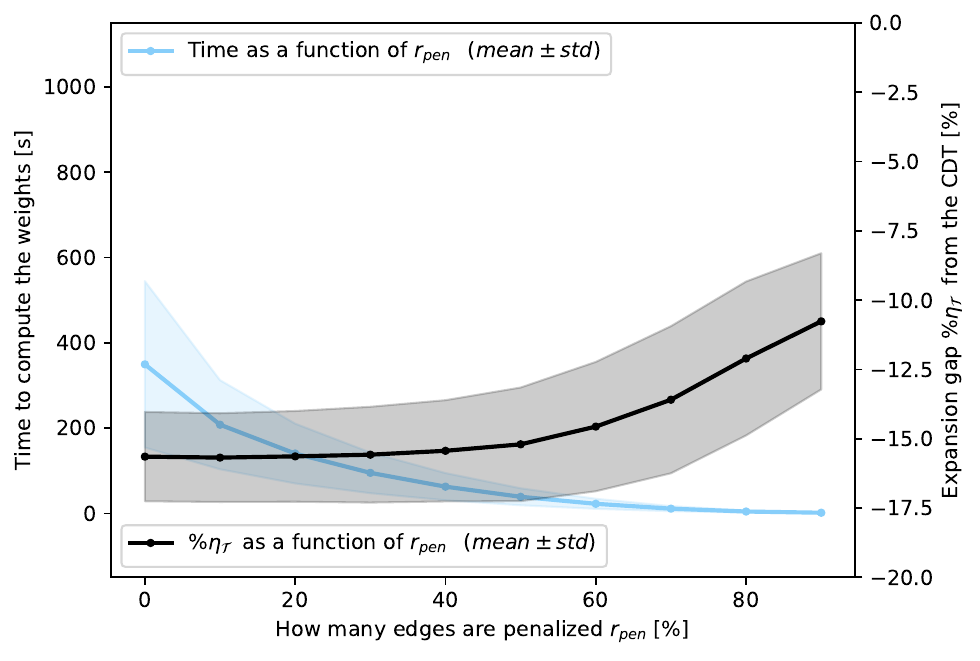}
		\caption{Tuning the parameter $r_{\mathit{pen}}$.}
		\label{fig:eval-weights}
	\end{subfigure}
	\caption{Tuning the parameters related to computing weights.}
\end{figure}

We start with the maximal sampling distance $d_{\mathit{samp}}$. 
A lower value means we get a better estimate of the MinVT weights, but it will take longer to compute them. 
To rule out the influence of $\eta_{\mathcal{T}}^h \neq \eta_{\mathcal{T}}$, we use simple polygon versions of the maps. 
The simple polygons are generated using the procedure in Alg.~\ref{alg:mwt} with $n_{\mathcal{P}} \;{=}\; \infty$.
See Fig.~\ref{fig:mesh-simple} for the example of a simple polygon generated from map \texttt{pl1} (Fig.~\ref{fig:mesh-holes}).
In the experiments, we use Alg.~\ref{alg:mwt-simple} to get the MinVT mesh for $d_{\mathit{samp}} \in \{ .125, .25, .5, 1, 2, 4, 8, 16, 32, 64, 128 \}$. 
Each is then evaluated on 1 million uniformly random query points and mean number of expansions $\eta_{\mathcal{T}}$ is recorded. 
Fig.~\ref{fig:eval-sample-dist} shows $\%\eta_{\mathcal{T}}$ (black, right axis) and computational time to compute the weights (light blue, left axis) as the function of $d_{\mathit{samp}}$.
$\%\eta_{\mathcal{T}}$ is the percentage gap from the CDT that we also evaluated for reference.
Both plots are averaged over the maps, and the standard deviation is included in the background as half-transparent. 
We can see that for $d_{\textit{samp}} < 1$ the value of $\%\eta_{\mathcal{T}}$ converges, but the time to get there is enormous.
On the other side for $d_{\textit{samp}} > 10$ we see very fast computations but poor quality.  
Values between $1$ and $10$ seem to provide the best compromise. 
We select $d_{\mathit{samp}} \in \{2,4,8\}$.

\begin{figure}[t]
	\centering
	\begin{subfigure}[b]{0.49\textwidth}
		\centering
		\includegraphics[width=1.0\textwidth]{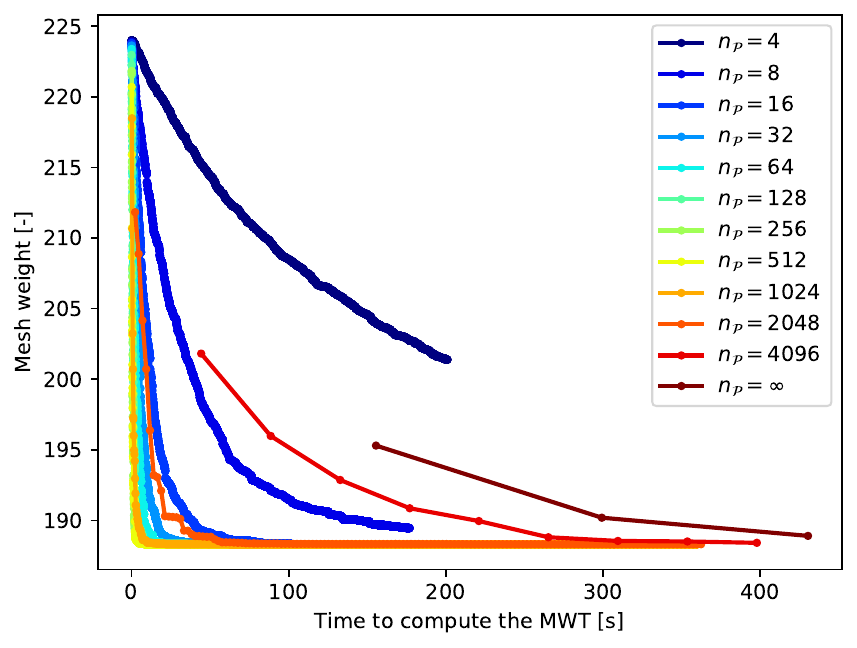}
	\end{subfigure}
	\hfill
	\begin{subfigure}[b]{0.49\textwidth}
		\centering
		\includegraphics[width=1.0\textwidth]{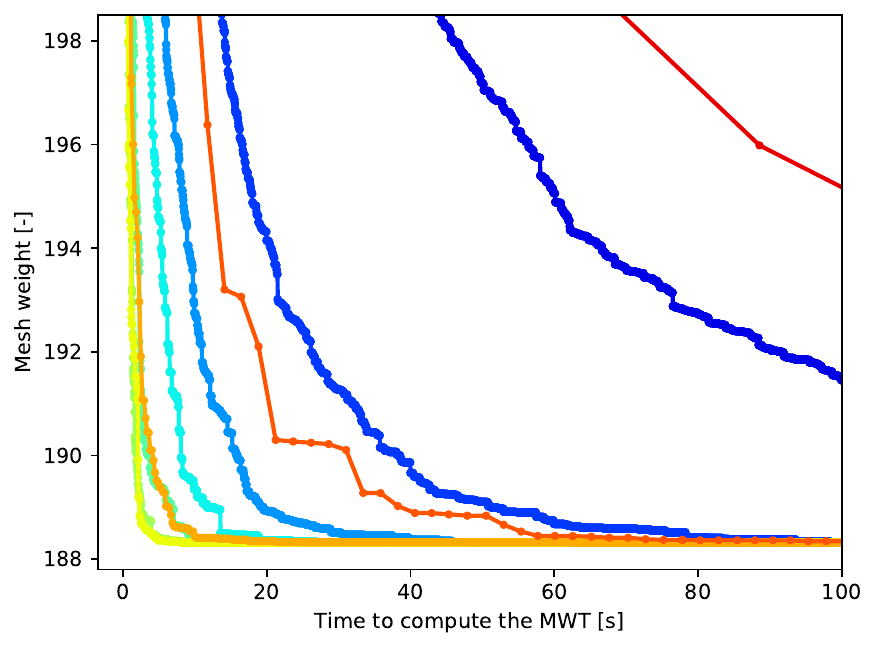}
	\end{subfigure}
	\caption{Tuning the MWT. The right figure is a detail of the left one.}
	\label{fig:eval-mwt}
\end{figure}
\begin{figure}[t]
	\centering
	\begin{subfigure}[b]{0.49\textwidth}
		\centering
		\includegraphics[width=1.0\textwidth]{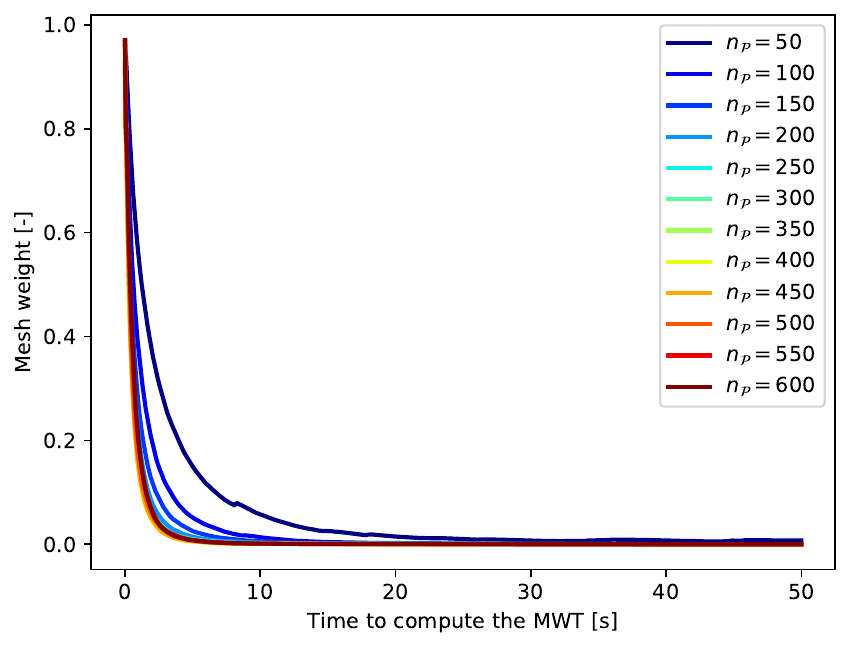}
	\end{subfigure}
	\hfill
	\begin{subfigure}[b]{0.49\textwidth}
		\centering
		\includegraphics[width=1.0\textwidth]{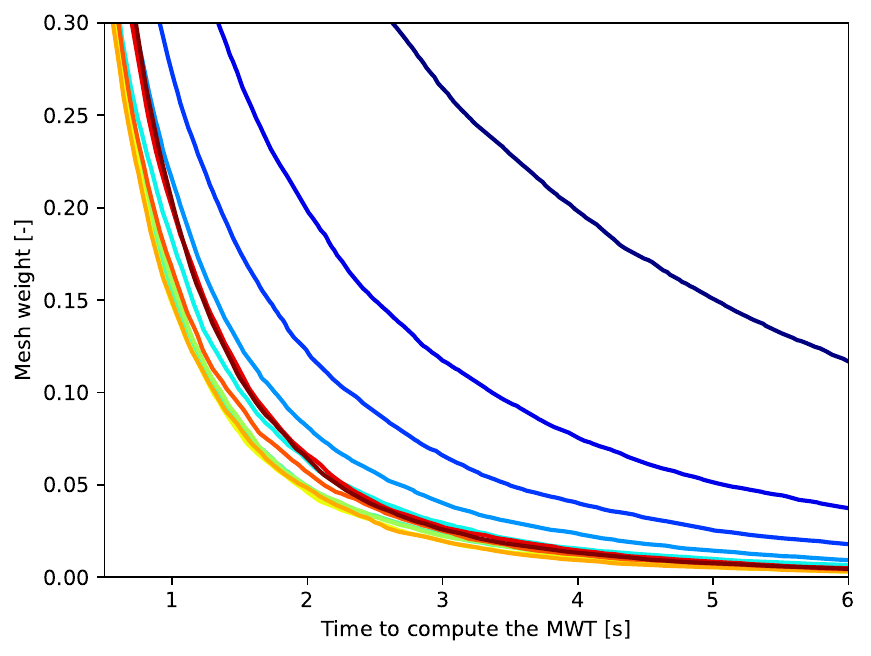}
	\end{subfigure}
	\caption{Tuning the MWT; higher resolution. The right figure is a detail.}
	\label{fig:eval-mwt-2}
\end{figure}

Next, we tune the parameters of Alg.~\ref{alg:mwt}: $n_\mathcal{P}$, $\mathit{it}_\mathit{max}$, and $t_\mathit{max}$. 
In the first set of experiments, we use $n_\mathcal{P} \in \big\{2^x\;\big\vert\;x\in\{2,\dots,12,\infty\}\big\}$, $\mathit{it}_\mathit{max} \;{=}\; 10000$, $t_\mathit{max} \;{=}\; 360$, and $d_{\mathit{samp}} \;{=}\; 2$.
We run Alg.~\ref{alg:mwt} several times with different settings on each map (with holes), and for each iteration, we record the current weight of the mesh and time from the start.
The mesh weight as a function of time is shown in Fig.~\ref{fig:eval-mwt}. 
For~clarity, we select to show results for just a single map (\texttt{pl4}) and we do not show the standard deviation.
For~$n_\mathcal{P} \in \{16,\dots,4096\}$, the weight converges within the chosen iteration and time window.
For~$n_\mathcal{P} \in \{64,\dots,1024\}$, it converges much more quickly compared to the other values. 
To select the best value, we are improving the resolution to $n_\mathcal{P} \in \{50,100,150,\dots,600\}$ and shrinking the time window to $t_\mathit{max} \;{=}\; 50$.
The corresponding plots are in Fig.~\ref{fig:eval-mwt-2}.
This time, we show the average plots over all maps.
Based on this second set of experiments, we are selecting $n_\mathcal{P} \;{=}\; 450$ as the best value because, at any given time, it provides the lowest weight and quickly converges. 
Based on the time and iteration window of the convergence we are selecting $\mathit{it}_\mathit{max} \;{=}\; 200$, and $t_\mathit{max} \;{=}\; 6$.

Finally, having $d_{\mathit{samp}} \;{=}\; 2$, $n_\mathcal{P} \;{=}\; 450$, $\mathit{it}_\mathit{max} \;{=}\; 200$, and $t_\mathit{max} \;{=}\; 6$, we~are still looking for a way to improve the time to compute the weights without a dramatic loss of quality. 
Based on the visual analysis of MinVT meshes, we have noticed that the mesh seldom chooses among the longest of the interior edges. 
An idea was born to penalize $r_\mathit{pen}\%$ of the longest interior edges $e_{i,j}$ by setting their weight to $w_{i,j} \;{=}\; 1000 \;{+}\; \vert e_{i,j} \vert$.
It saves time because determining the penalized weight is significantly less computationally demanding than the non-penalized weight $w_{i,j} \;{=}\; \mathrm{Area}\big(\mathcal{V}_e(e_{i,j}))$.
The~penalization means that the $(100\;{-}\;r_\mathit{pen})\%$ of the shortest edges are always preferred over the $r_\mathit{pen}\%$ of the longest edges.
If~there is no suitable short edge to be selected, then MWT at least optimizes among the long edges according to their length. 
We run the same set of experiments as when tuning $d_{\mathit{samp}}$ except the variable parameter is now $r_\mathit{pen}$, and the maps are with holes. 
The~corresponding plots are shown in~Fig.~\ref{fig:eval-weights}. 
We~select $r_\mathit{pen}\;{=}\;0$ as the default and $r_\mathit{pen}\;{=}\;50$ as the best compromise between quality and time. 

\subsection{Mesh Evaluation: TEA}
\label{sec:eval-tea}

\begin{sidewaystable}
	\sidewaystablefn%
	\begin{center}
		\begin{minipage}{\textheight}
			\setlength{\tabcolsep}{3pt}
			\caption{Performance of the TEA with various triangular meshes.}
			\label{tab:res-tea}
			\begin{tiny}
				\begin{tabular*}{\textheight}{@{\extracolsep{\fill}}l*{29}{r}@{\extracolsep{\fill}}}
					\toprule
					& \multicolumn{2}{r}{CDT} & \multicolumn{3}{r}{MinLT} & \multicolumn{3}{r}{MinVT-2} & \multicolumn{3}{r}{MinVT-4} & \multicolumn{3}{r}{MinVT-8} & \multicolumn{3}{r}{MinVTO-2} & \multicolumn{3}{r}{MinVTO-4} & \multicolumn{3}{r}{MinVTO-8} & \multicolumn{3}{r}{MaxLT} & \multicolumn{3}{r}{MaxVT-2} \\
					\cmidrule(l){2-3} \cmidrule(l){4-6} \cmidrule(l){7-9} \cmidrule(l){10-12} \cmidrule(l){13-15} \cmidrule(l){16-18} \cmidrule(l){19-21} \cmidrule(l){22-24} \cmidrule(l){25-27} \cmidrule(l){28-30}
					map & $\eta_\mathcal{T}$ & $\bar{t}_q$ & $t_c$ & $\%\eta_\mathcal{T}$ & $\%\bar{t}_q$ & $t_c$ & $\%\eta_\mathcal{T}$ & $\%\bar{t}_q$ & $t_c$ & $\%\eta_\mathcal{T}$ & $\%\bar{t}_q$ & $t_c$ & $\%\eta_\mathcal{T}$ & $\%\bar{t}_q$ & $t_c$ & $\%\eta_\mathcal{T}$ & $\%\bar{t}_q$ & $t_c$ & $\%\eta_\mathcal{T}$ & $\%\bar{t}_q$ & $t_c$ & $\%\eta_\mathcal{T}$ & $\%\bar{t}_q$ & $t_c$ & $\%\eta_\mathcal{T}$ & $\%\bar{t}_q$ & $t_c$ & $\%\eta_\mathcal{T}$ & $\%\bar{t}_q$ \\
					\midrule
					\tt 2p1 & 228.5 & 10.33 & 5 & -9 & -12 & 123 & -12 & -14 & 56 & -11 & -12 & 26 & -6 & -9 & 17 & -12 & -14 & 10 & -10 & -12 & 7 & -6 & -10 & 4 & 97 & 47 & 122 & 106 & 51 \\
					\tt 2p2 & 212.8 & 9.31 & 4 & -11 & -13 & 93 & -16 & -16 & 43 & -14 & -16 & 20 & -9 & -11 & 16 & -15 & -15 & 9 & -13 & -14 & 6 & -9 & -12 & 4 & 99 & 48 & 93 & 106 & 50 \\
					\tt 2p4 & 83.1 & 4.01 & 4 & -7 & -11 & 21 & -15 & -16 & 12 & -14 & -14 & 7 & -11 & -14 & 6 & -14 & -14 & 5 & -13 & -14 & 4 & -10 & -13 & 4 & 56 & 18 & 21 & 78 & 30 \\
					\tt 4p1 & 232.1 & 10.93 & 5 & -8 & -12 & 203 & -13 & -16 & 91 & -12 & -14 & 41 & -7 & -11 & 30 & -13 & -15 & 15 & -11 & -13 & 10 & -6 & -11 & 5 & 95 & 44 & 203 & 109 & 52 \\
					\tt 4p3 & 177.3 & 9.15 & 9 & -7 & -12 & 220 & -13 & -16 & 100 & -11 & -15 & 47 & -8 & -12 & 28 & -12 & -15 & 17 & -10 & -14 & 12 & -7 & -12 & 9 & 79 & 30 & 219 & 94 & 41 \\
					\midrule
					\tt 6p1 & 250.8 & 11.68 & 6 & -10 & -14 & 260 & -16 & -18 & 118 & -15 & -17 & 53 & -12 & -15 & 35 & -16 & -18 & 18 & -14 & -17 & 11 & -12 & -15 & 6 & 89 & 40 & 259 & 107 & 48 \\
					\tt 6p3 & 184.1 & 8.71 & 6 & -12 & -14 & 200 & -18 & -19 & 92 & -17 & -18 & 43 & -14 & -16 & 26 & -18 & -18 & 14 & -16 & -17 & 9 & -14 & -15 & 5 & 90 & 38 & 199 & 109 & 48 \\
					\tt ch2 & 431.0 & 17.38 & 4 & -15 & -17 & 1368 & -21 & -21 & 602 & -20 & -20 & 266 & -18 & -19 & 43 & -20 & -19 & 21 & -19 & -20 & 11 & -17 & -18 & 4 & 128 & 65 & 1367 & 144 & 71 \\
					\tt ch3 & 185.7 & 9.09 & 6 & -9 & -13 & 251 & -15 & -17 & 113 & -13 & -16 & 51 & -9 & -13 & 30 & -14 & -17 & 16 & -12 & -16 & 10 & -9 & -13 & 6 & 96 & 42 & 251 & 104 & 47 \\
					\tt ch4 & 188.2 & 9.37 & 8 & -6 & -11 & 225 & -11 & -14 & 101 & -9 & -13 & 46 & -4 & -9 & 27 & -11 & -14 & 15 & -9 & -13 & 11 & -4 & -10 & 7 & 80 & 35 & 224 & 107 & 50 \\
					\midrule
					\tt end & 201.1 & 10.23 & 10 & -7 & -12 & 521 & -15 & -17 & 233 & -13 & -17 & 107 & -10 & -14 & 44 & -14 & -16 & 24 & -13 & -16 & 16 & -9 & -14 & 9 & 85 & 35 & 521 & 105 & 49 \\
					\tt pl1 & 95.7 & 4.51 & 3 & -4 & -7 & 8 & -8 & -9 & 5 & -6 & -8 & 4 & -0 & -5 & 4 & -7 & -9 & 3 & -6 & -8 & 3 & 0 & -4 & 2 & 68 & 30 & 7 & 73 & 32 \\
					\tt pl2 & 126.1 & 6.39 & 6 & -10 & -14 & 91 & -15 & -18 & 43 & -14 & -17 & 21 & -10 & -14 & 17 & -15 & -17 & 11 & -13 & -16 & 8 & -9 & -13 & 6 & 68 & 26 & 90 & 82 & 31 \\
					\tt pl3 & 248.6 & 12.42 & 8 & -10 & -15 & 406 & -15 & -18 & 180 & -13 & -18 & 79 & -9 & -15 & 56 & -14 & -19 & 28 & -13 & -17 & 17 & -8 & -15 & 8 & 104 & 45 & 406 & 114 & 49 \\
					\tt pl5 & 164.1 & 7.75 & 5 & -7 & -10 & 144 & -14 & -15 & 65 & -12 & -13 & 30 & -8 & -11 & 18 & -13 & -13 & 10 & -12 & -13 & 7 & -8 & -11 & 4 & 76 & 34 & 143 & 93 & 43 \\
					\midrule
					\tt rs1 & 143.9 & 6.78 & 4 & -7 & -9 & 57 & -11 & -12 & 27 & -8 & -11 & 13 & -2 & -7 & 9 & -10 & -12 & 6 & -8 & -10 & 5 & -2 & -7 & 4 & 79 & 34 & 56 & 86 & 39 \\
					\tt rs2 & 87.2 & 4.24 & 4 & -4 & -8 & 16 & -10 & -12 & 9 & -8 & -9 & 6 & -4 & -8 & 5 & -8 & -10 & 4 & -6 & -9 & 4 & -3 & -7 & 3 & 70 & 28 & 16 & 82 & 34 \\
					\tt rs4 & 180.3 & 8.83 & 7 & -9 & -14 & 159 & -14 & -17 & 73 & -12 & -15 & 34 & -8 & -13 & 22 & -13 & -16 & 13 & -12 & -15 & 9 & -7 & -12 & 6 & 91 & 38 & 159 & 99 & 42 \\
					\tt rs6 & 160.0 & 8.27 & 10 & -8 & -13 & 171 & -15 & -18 & 80 & -13 & -17 & 39 & -8 & -13 & 26 & -13 & -17 & 17 & -11 & -16 & 13 & -7 & -13 & 9 & 66 & 25 & 171 & 88 & 39 \\
					\tt rs7 & 167.3 & 7.92 & 5 & -7 & -11 & 131 & -12 & -15 & 61 & -11 & -13 & 28 & -7 & -12 & 19 & -12 & -15 & 11 & -11 & -14 & 7 & -7 & -11 & 4 & 82 & 35 & 130 & 102 & 45 \\
					\midrule
					\tt sx1 & 144.0 & 6.47 & 4 & -7 & -10 & 85 & -13 & -13 & 40 & -11 & -12 & 19 & -6 & -10 & 13 & -12 & -13 & 8 & -10 & -12 & 6 & -6 & -9 & 3 & 80 & 36 & 85 & 104 & 49 \\
					\tt sx3 & 142.5 & 7.42 & 8 & -7 & -13 & 182 & -13 & -16 & 84 & -11 & -16 & 40 & -8 & -14 & 29 & -12 & -17 & 17 & -11 & -17 & 12 & -7 & -13 & 8 & 78 & 28 & 181 & 95 & 38 \\
					\tt sx4 & 128.8 & 6.68 & 8 & -7 & -13 & 125 & -13 & -17 & 59 & -12 & -16 & 29 & -7 & -14 & 21 & -12 & -17 & 13 & -10 & -15 & 10 & -6 & -11 & 8 & 73 & 28 & 125 & 92 & 36 \\
					\tt sx5 & 79.1 & 3.79 & 4 & -8 & -10 & 19 & -16 & -14 & 10 & -15 & -15 & 6 & -13 & -13 & 6 & -13 & -13 & 4 & -13 & -8 & 4 & -11 & -12 & 3 & 43 & 13 & 19 & 63 & 24 \\
					\tt sx6 & 194.1 & 9.46 & 5 & -8 & -14 & 224 & -14 & -18 & 102 & -12 & -16 & 47 & -9 & -14 & 33 & -14 & -17 & 17 & -12 & -16 & 10 & -9 & -14 & 5 & 85 & 34 & 224 & 102 & 42 \\
					\midrule
					\bf avg & \bf 177.5 & \bf 8.44 & \bf 6 & \bf -8 & \bf -12 & \bf 212 & \bf -14 & \bf -16 & \bf 96 & \bf -12 & \bf -15 & \bf 44 & \bf -8 & \bf -12 & \bf 23 & \bf -13 & \bf -15 & \bf 13 & \bf -12 & \bf -14 & \bf 9 & \bf -8 & \bf -12 & \bf 5 & \bf 82 & \bf 35 & \bf 212 & \bf 98 & \bf 43 \\
					\cmidrule(l){2-3} \cmidrule(l){4-6} \cmidrule(l){7-9} \cmidrule(l){10-12} \cmidrule(l){13-15} \cmidrule(l){16-18} \cmidrule(l){19-21} \cmidrule(l){22-24} \cmidrule(l){25-27} \cmidrule(l){28-30}
					& \multicolumn{2}{r}{CDT} & \multicolumn{3}{r}{MinLT} & \multicolumn{3}{r}{MinVT-2} & \multicolumn{3}{r}{MinVT-4} & \multicolumn{3}{r}{MinVT-8} & \multicolumn{3}{r}{MinVTO-2} & \multicolumn{3}{r}{MinVTO-4} & \multicolumn{3}{r}{MinVTO-8} & \multicolumn{3}{r}{MaxLT} & \multicolumn{3}{r}{MaxVT-2} \\
					\botrule
				\end{tabular*}
			\end{tiny}
			\footnotetext{
				\emph{Legend:} $\eta_\mathcal{T}$ is the mean number of edge expansions; $\bar{t}_q$ is the mean query computational time; $t_c$ is the construction time of the mesh; $\%\mathit{val}$ means a relative percentage gap of $\mathit{val}$ from the CDT value ($\mathit{val}_\mathrm{CDT}$).
				Every value of $\eta_\mathcal{T}$ and $\bar{t}_q$ is based on one million queries; the values in the \textbf{avg} row are averaged over all test maps.
			}
			\footnotetext{
				\emph{Units:} $\eta_\mathcal{T}$ is a count (the unit is 1); $\bar{t}_q$ is in microseconds, $t_c$ is in seconds; $\%\eta_\mathcal{T}$, and $\%\bar{t}_q$ are in percents.
			}
		\end{minipage}
	\end{center}
\end{sidewaystable}

The final evaluation is done on the 25 map instances from~\cite{Harabor2022}, which are not marked in Tab.~\ref{tab:maps}, in the following way.
First, the mesh under evaluation is constructed to represent the given map. 
The CDT is constructed using Triangle~\cite{Shewchuk1996}, all the other meshes using the MWT (Alg.~\ref{alg:mwt}).
The time of construction $t_c$ is recorded.  
Then, one million uniformly random query points are generated inside the map using a pseudo-random generator initialized always with the same seed to ensure fair conditions for every mesh. 
Then, the TEA is run to compute a visibility region for each query point. 
The number of edge expansions $\eta_q$ done by the TEA and computational time $t_q$ are recorded in the process.
These values are then averaged to obtain $\eta_\mathcal{T}$ and $\bar{t}_q$. 
We report the metric's absolute values for the CDT, and for other meshes, we report the relative percentage gap from the reference CDT value according to~\eqref{eq:gap}.

Besides CDT as the reference, we evaluate a number of MinVT variants, all of which can be enumerated as $\{\text{`'},\text{d-}\}$MinVT$\{\text{`'},\text{O}\}$-$\{\text{2},\text{4},\text{8}\}$. 
The~prefix d- means limited visibility range $d$ was used when computing the weights. 
The~variant without the prefix means $d=\infty$.
The O stands for \emph{Optimized}, and it refers to $r_{\mathit{pen}} \;{=}\; 50$. 
No O implies $r_{\mathit{pen}} \;{=}\; 0$.
The suffix number is the value~of~$d_{\mathit{samp}}$. 
We also evaluate the \emph{minimum length triangulation} MinLT with weights defined as $w_{i,j} \;{=}\; \vert e_{i,j} \vert$.
Furthermore, we think evaluating the full range of meshes is interesting, not just the good ones and the reference. 
Therefore, we also include the MaxLT and MaxVT-2, which have negative weights of MinLT and MinVT-2, respectively. 
We show all eventuated meshes constructed on a unique map\footnote{
	This map was specifically designed to emphasize the differences between the meshes.
} in Fig.~\ref{fig:meshes}. 

The complete results for every test map and mesh type, excluding MinVT with the d- prefix, are shown in Tab.~\ref{tab:res-tea}.
We can see that the CDT performs 177.5 expansions and runs for 8.44 $\mu$s per query, on average over all maps. 
The~CDT construction time is effectively zero compared to other meshes. 
The~average MinLT takes $6\,\text{sec}$ to construct and improves the number of expansions and query time by $8\,\%$ and $12\,\%$, respectively.
The best mesh in terms of query performance is MinVT-2, improving on CDT by $14\,\%$ and $16\,\%$ on average, according to the two respective metrics.
This observation is essential because MinVT-2 is the closest approximation of the proposed optimal mesh, thus validating the overall approach. 

Constructing the MinVT-2 takes $212\,\text{sec}$ on average.
We can observe, however, that the MinVT construction times greatly depend on parameters $d_\mathit{samp}$ and $r_{\mathit{pen}}$, which affect how the weights are computed. 
This observation and the fact that the MWT has a maximal run time limit set to $6\,\text{sec}$ suggest that computing the weights is the main bottleneck.
It could be addressed by more efficient computation of the region visible from an edge, for example, with a specialized algorithm based on the TEA that does not require edge sampling. 
We~are halfway through implementing this algorithm, but it could not yet be used in this study due to insufficient robustness.
We~expect this novel algorithm to better approximate the optimal mesh than the MinVT-2 and approach the MinLT with the construction times. 
In the current state of affairs, however, it seems that MinVTO-4 provides the best compromise between construction and query times among the MinVT meshes with $13\,\text{sec}$ of construction, and $12$/$14\%$ average improvement over the CDT.

Lastly, note that using the MaxLT and MaxVT-2 meshes results in the worst query performance by far, according to expectations. 

\subsection{Mesh Evaluation: d-TEA}
\label{sec:eval-dtea} 

\begin{sidewaystable}
	\sidewaystablefn%
	\begin{center}
		\begin{minipage}{\textheight}
			\setlength{\tabcolsep}{3pt}
			\caption{Performance of the d-TEA with various triangular meshes for various values of $d$ (summary).}
			\label{tab:res-dtea}
			\begin{tiny}
				\begin{tabular*}{\textheight}{@{\extracolsep{\fill}}l*{24}{r}@{\extracolsep{\fill}}}
					\toprule
					& \multicolumn{3}{r}{$d\,{=}\,\infty$} & \multicolumn{3}{r}{$d\,{=}\,128$} & \multicolumn{3}{r}{$d\,{=}\,64$} & \multicolumn{3}{r}{$d\,{=}\,32$} & \multicolumn{3}{r}{$d\,{=}\,16$} & \multicolumn{3}{r}{$d\,{=}\,8$} & \multicolumn{3}{r}{$d\,{=}\,4$} & \multicolumn{3}{r}{$d\,{=}\,2$} \\
					\cmidrule(l){2-4} \cmidrule(l){5-7} \cmidrule(l){8-10} \cmidrule(l){11-13} \cmidrule(l){14-16} \cmidrule(l){17-19} \cmidrule(l){20-22} \cmidrule(l){23-25} 
					mesh type & $\bar{t}_c$ & $\%\bar{\eta}_\mathcal{T}$ & $\%\bar{t}_q$ & $\bar{t}_c$ & $\%\bar{\eta}_\mathcal{T}$ & $\%\bar{t}_q$ & $\bar{t}_c$ & $\%\bar{\eta}_\mathcal{T}$ & $\%\bar{t}_q$ & $\bar{t}_c$ & $\%\bar{\eta}_\mathcal{T}$ & $\%\bar{t}_q$ & $\bar{t}_c$ & $\%\bar{\eta}_\mathcal{T}$ & $\%\bar{t}_q$ & $\bar{t}_c$ & $\%\bar{\eta}_\mathcal{T}$ & $\%\bar{t}_q$ & $\bar{t}_c$ & $\%\bar{\eta}_\mathcal{T}$ & $\%\bar{t}_q$ & $\bar{t}_c$ & $\%\bar{\eta}_\mathcal{T}$ & $\%\bar{t}_q$ \\
					\midrule
					CDT\footnotemark[1] & 0 & 177.5\textsuperscript{$\star$} & 8.44\textsuperscript{$\star$} & 0 & 155.8 & 10.14 & 0 & 111.4 & 7.85 & 0 & 59.6 & 4.90 & 0 & 26.1 & 2.74 & 0 & 10.6 & 1.56 & 0 & 4.5 & 0.99 & 0 & 2.1 & 0.73 \\
					\midrule
					MinLT & 6\textsuperscript{$\star$} & -8.1\textsuperscript{$\star$} & -12.1\textsuperscript{$\star$} & 6 & -7.8 & -10.6 & 6 & -6.9 & -8.6 & 6 & -6.4 & -5.9 & 6 & -6.9 & -2.4 & 6 & -9.3 & 0.2 & 6 & -12.3 & 1.8 & 6 & -14.3 & 3.1 \\
					\midrule
					MinVT-2 & 212\textsuperscript{$\star$} & -13.9\textsuperscript{$\star$} & -15.9\textsuperscript{$\star$} & 212 & -12.4 & -12.9 & 212 & -8.9 & -8.8 & 212 & -5.2 & -2.9 & 212 & -2.2 & 3.4 & 212 & -0.7 & 7.2 & 212 & -0.4 & 8.8 & 212 & 0.0 & 9.0 \\
					MinVT-4 & 96\textsuperscript{$\star$} & -12.3\textsuperscript{$\star$} & -14.7\textsuperscript{$\star$} & 96 & -11.0 & -12.1 & 96 & -8.0 & -8.4 & 96 & -4.8 & -3.0 & 96 & -2.3 & 2.8 & 96 & -1.4 & 6.1 & 96 & -1.3 & 7.7 & 96 & -0.9 & 8.2 \\
					MinVT-8 & 44\textsuperscript{$\star$} & -8.2\textsuperscript{$\star$} & -12.3\textsuperscript{$\star$} & 44 & -7.2 & -10.1 & 44 & -4.8 & -7.0 & 44 & -2.3 & -1.8 & 44 & -0.3 & 3.3 & 44 & 0.7 & 6.8 & 44 & 2.0 & 9.1 & 44 & 3.7 & 9.7 \\
					\midrule
					MinVTO-2 & 23\textsuperscript{$\star$} & -13.1\textsuperscript{$\star$} & -15.3\textsuperscript{$\star$} & 23 & -11.9 & -12.7 & 23 & -9.2 & -9.4 & 23 & -6.3 & -4.4 & 23 & -4.2 & 1.0 & 23 & -3.8 & 4.5 & 23 & -4.4 & 6.2 & 23 & -4.7 & 6.8 \\
					MinVTO-4 & 13\textsuperscript{$\star$} & -11.5\textsuperscript{$\star$} & -14.1\textsuperscript{$\star$} & 13 & -10.5 & -11.9 & 13 & -8.1 & -8.8 & 13 & -5.6 & -4.3 & 13 & -4.0 & 0.6 & 13 & -4.1 & 3.9 & 13 & -4.8 & 5.5 & 13 & -4.9 & 6.1 \\
					MinVTO-8 & 9\textsuperscript{$\star$} & -7.6\textsuperscript{$\star$} & -11.8\textsuperscript{$\star$} & 9 & -6.8 & -9.8 & 9 & -4.8 & -7.1 & 9 & -2.9 & -2.9 & 9 & -1.6 & 1.6 & 9 & -1.3 & 4.8 & 9 & -0.7 & 6.7 & 9 & 0.6 & 7.7 \\
					\midrule
					d-MinVT-2 & - & - & -  & 189 & -12.7 & -13.1 & 131 & -10.4 & -10.5 & 71 & -8.8 & -6.6 & 35 & -8.8 & -2.9 & 20 & -10.7 & -0.0 & 15 & -13.2 & 2.4 & 13 & -15.1 & 2.6 \\
					d-MinVT-4 & - & - & -  & 93 & -11.2 & -12.3 & 66 & -9.2 & -9.7 & 39 & -8.0 & -6.3 & 22 & -8.2 & -2.7 & 14 & -10.2 & 0.5 & 11 & -12.8 & 2.2 & 10 & -14.3 & 2.9 \\
					d-MinVT-8 & - & - & -  & 47 & -7.5 & -10.1 & 35 & -6.1 & -7.8 & 23 & -5.7 & -4.9 & 15 & -6.6 & -1.6 & 11 & -8.6 & 1.1 & 9 & -10.5 & 3.9 & 9 & -12.0 & 4.2 \\
					\midrule
					d-MinVTO-2 & - & - & -  & 27 & -12.1 & -13.0 & 24 & -10.0 & -10.0 & 18 & -8.6 & -6.6 & 13 & -8.6 & -2.6 & 10 & -10.5 & 0.2 & 9 & -13.0 & 2.9 & 8 & -14.9 & 3.0 \\
					d-MinVTO-4 & - & - & -  & 16 & -10.6 & -11.9 & 15 & -8.9 & -9.6 & 12 & -7.8 & -6.0 & 10 & -8.0 & -2.5 & 8 & -10.0 & 0.5 & 8 & -12.6 & 2.4 & 7 & -14.1 & 3.2 \\
					d-MinVTO-8 & - & - & -  & 12 & -7.0 & -9.7 & 11 & -5.9 & -7.8 & 10 & -5.5 & -5.0 & 9 & -6.4 & -1.4 & 8 & -8.5 & 1.1 & 7 & -10.3 & 3.8 & 7 & -11.9 & 3.7 \\
					\cmidrule(l){2-4} \cmidrule(l){5-7} \cmidrule(l){8-10} \cmidrule(l){11-13} \cmidrule(l){14-16} \cmidrule(l){17-19} \cmidrule(l){20-22} \cmidrule(l){23-25} 
					& \multicolumn{3}{r}{$d\,{=}\,\infty$} & \multicolumn{3}{r}{$d\,{=}\,128$} & \multicolumn{3}{r}{$d\,{=}\,64$} & \multicolumn{3}{r}{$d\,{=}\,32$} & \multicolumn{3}{r}{$d\,{=}\,16$} & \multicolumn{3}{r}{$d\,{=}\,8$} & \multicolumn{3}{r}{$d\,{=}\,4$} & \multicolumn{3}{r}{$d\,{=}\,2$} \\
					\botrule
				\end{tabular*}
			\end{tiny}
			\footnotetext{
				\emph{Legend:} $d$ is the visibility range. Otherwise, the legend is the same as in Tab.~\ref{tab:res-tea}, but all the values are averaged over all test maps.
			}
			\footnotetext{
				\emph{Units:} $d$ is in meters; $\bar{t}_c$ is in seconds; $\%\bar{\eta}_\mathcal{T}$, and $\%\bar{t}_q$ are in percents except for CDT (see footnote\footnotemark[1]).
			}
			\footnotetext{
				\emph{Note:} The values marked with \textsuperscript{$\star$} also appear at the bottom of Tab.~\ref{tab:res-tea} (\textbf{avg} row). 
			}
			\footnotetext[1]{
				The values in $\%\bar{\eta}_\mathcal{T}$ and $\%\bar{t}_q$ columns are absolute for the CDT: $\bar{\eta}_\mathcal{T}$ is a mean count, $\bar{t}_q$ is in microseconds.
				The same values are relative percentage gaps w.r.t. CDT for the other mesh types.
			}
		\end{minipage}
	\end{center}
\end{sidewaystable}

The previous results in Tab.~\ref{tab:res-dtea} can be seen as the results for the d-TEA with unlimited visibility range. 
In this section, we present the d-TEA results for some other, finite, values of $d$: $128$, $64$, $32$, $16$, $8$, $4$, and $2$.\footnote{
	Recall that we measure $d$ in meters, which is arbitrary in essence. 
    What matters is that we measure it in the same unit as the maps and parameters like $d_{\mathit{samp}}$. 
    Readers can compare the $d$ values used in our experiments with the dimensions of the maps shown in Table~\ref{tab:maps}.
} 
Recall that for finite values, the output region returned by d-TEA is intersected with a circle of radius $d$ centered in the query point. 
We note that we measure the computational time of this operation, $t_\circ$, and add it to the total query time $t_q$.

The extended results are shown in Tab.~\ref{tab:res-dtea}.
We present only the averaged values (over all maps) because we could not fit that many tables in this paper. 
Note that the whole Tab.~\ref{tab:res-tea} is virtually contained in Tab.~\ref{tab:res-dtea} (see the ${}^\star$-marked values). 
Also, be aware that the CDT values are still absolute, while for the other meshes, we present relative values except $\bar{t}_c$. 
They extended results newly include the d-MinVT meshes for the finite values of $d$.

We can see that for every mesh, the construction time, number of expansions, and query times are all getting lower with lower values of $d$ due to the early exit strategy of d-TEA.
However, this rule has one exception: the query times for $d\;{=}\;128$ are higher than for $d=\infty$. 
The numbers of expansions are relatively similar for these two specific values, but for $d\;{=}\;128$, there is an additional computational burden in resolving the circle intersection.
Time $t_\circ$ to compute the circle intersection outweighs the time saved by sparing some of the expansions; thus, there is an increase in total query time. 

Let us now compare the meshes.
We can see that across all finite values of~$d$, the d-MinVT-2 mesh consistently provides the least number of expansions, while MinVT-2 is often worse than the MinLT.
This confirms that the optimal mesh, as we defined it, is not generally the same for the d-TEA as for the~TEA.
Furthermore, according to the expansions, the gap between d-MinVT-2 and MinLT is getting lower with lower visibility ranges. 
This is caused by the fact that their weights are getting more  (approximately) proportional. Note that as we lower the visibility range, the area visible from a segment depends more proportionally on the segment's length. 

The last observation we want to point out is the increasing inconsistency between the numbers of expansions and query times for decreasing $d$. 
Starting with $d = 8$, the CDT is, in fact, the best-performing mesh according to the measured query time. 
This is contrary to what the expansions tell us.
We~explain this the following way. 
Unlike the TEA, the d-TEA outputs different (temporary) regions for the same queries and various meshes. 
The circle intersection only then unifies these regions to be the same.
The temporary regions' difference is due to d-TEA expanding until full edges are outside the visibility range, which depends on how the edges are organized. 
Different temporary regions affect the circle intersection computation times, which may outweigh the effect of lowering the number of expansions. 
Therefore, this whole observation suggests that when the visibility range is considerably limited, the CDT might be closer to the actual optimal mesh than the proposed mesh minimizing the number of expansions. 
But what is the actual optimal mesh, then?
Probably the one that minimizes the combination of the number of expansions and the laboriousness of the circle intersection. 
Formalizing this idea is an interesting open question. 

\section{Conclusion}
\label{sec:conclusion}

In this paper and its proceedings' predecessor~\cite{Mikula2022b}, we addressed the problem of improving the query performance of the TEA~\cite{Bungiu2014} for computing visibility regions with only one variable component---the instance of the triangular mesh, i.e., the preprocessing structure.
Based on the equations in Sec.~\ref{sec:problem}, experiments on the 11 maps from variuous sources performed in~\cite{Mikula2022b}, and more thorough experiments on the challenging 25 maps~\cite{Harabor2022} from the Iron Harvest videogame achieved in this paper, we can now answer the following scientific and practical questions.

\begin{itemize}
	\item \emph{Does the choice of mesh influence the query performance of the algorithm?}---Yes. We measured the mean query time of 1 million identical queries for many different meshes. 
	The largest obtained gap between two meshes was $116\%$; the better of the two had $13.7\,\mu s$, and the worse $29.7\,\mu s$.
	\item \emph{Can some other more exact metric be used to access the mesh performance?}---Yes. The number of triangle edge expansions is approximately proportional to the measured computational time. 
	\item \emph{Is there an optimal mesh?}---The optimal mesh can be defined as in Eq.~\eqref{eq:Tstar} based on the expected number of edge expansions assuming the query points are drawn from a known probability distribution. 
	The proposed MinVT defined in Eq.~\eqref{eq:Tstarfinal} well approximates the optimal mesh, assuming all possible queries are equally likely and the map resembles a real-world environment.
	\item \emph{Is there a way to construct the optimal mesh?}---The problem of finding the optimal mesh is NP-hard. 
	However, in Sec.~\ref{sec:solution} we propose a heuristic algorithm that provides good approximations; we include a guide to tuning its parameters and describe the implementation, which is publicly available.
	\item \emph{What is the expected improvement and what are the costs?}---The following values are the average results from our experiments. 
	The~MinVT shortens the mean query times by about $12$-$16\%$ for the cost of $9$-$212\,\text{sec}$ of preprocessing, depending on the parameters, compared to the reference CDT. 
	The~MinLT, i.e., the mesh minimizing edge lengths, also performs better than CDT by $12\%$ and takes $6\,\text{sec}$ to construct. 
	The~CDT, on the other hand, takes effectively zero time to construct. 
	\item \emph{What if the visibility range is limited?}---The d-TEA, which is part of our implementation, can handle it efficiently. 
	Depending on the range, the best-performing mesh is either d-MinVT or CDT. 
	The CDT is better for small ranges because it outputs temporary regions which are easier to intersect with a circle. 
	The optimal d-TEA mesh would need to include this operation in the optimizing criterion. 
\end{itemize}

In conclusion, our approach is well-suited for enhancing the performance of offline applications that involve computing millions of queries in complex real-world environments, where extended preprocessing time is not a concern. 
Otherwise, the CDT is the best choice for its simplicity and fast construction.
The ideas presented in this paper may inspire efforts to improve other mesh algorithms in a similar manner. 
Furthermore, our implementation is available to benefit the community.

\section*{Acknowledgments}

This work was co-funded by the European Union under the project Robotics and advanced industrial production (reg. no. CZ.02.01.01/00/22\_008/0004590) and by the Grant Agency of the Czech Technical University in Prague, grant no. SGS23/175/OHK3/3T/13.



\section*{Declarations}

\textbf{Conflict of interest} On behalf of all authors, the corresponding author states that there is no conflict of interest.

\begin{appendices}
	
	\section{Map Properties}
	\label{sec:maps}
	In this paper, we use as the benchmark instances the 35 polygonal maps from~\cite{Harabor2022}, which are based on the videogame Iron Harvest developed by KING Art Games.
    The instances were initially designed for the pathfinding community but also serve as optimal benchmarks for evaluating algorithms that compute visibility regions. 
    The authors offer three types of representations: polygonal, mesh, and grid. 
    Given that our paper focuses on generating meshes from polygons, we have opted for the polygonal representation. 
    To obtain the single polygon with holes essential for our algorithms, we specifically choose the largest polygon from the provided representation and incorporate all the holes it encompasses.
	
		\begin{table}[t]
		\begin{center}
			\begin{minipage}{\textwidth}
				\setlength{\tabcolsep}{3pt}
				\caption{Properties of the polygonal maps.}\label{tab:maps}
				\begin{tabular*}{\textwidth}{@{\extracolsep{\fill}}lc*{5}{r}c@{\extracolsep{\fill}}}
					\toprule%
					Full name & Abbrev. & $n$ [1] & $h$ [1] & $x$ [m] & $y$ [m] & $a$ [m\textsuperscript{2}] & Used in tuning? \\
					\midrule 
					\tt scene\_mp\_2p\_01 & \tt 2p1 & 3219 & 263 & 200.00 & 210.00 & 35095.74 & \\
					\tt scene\_mp\_2p\_02 & \tt 2p2 & 2318 & 165 & 270.00 & 270.00 & 54034.51 & \\
					\tt scene\_mp\_2p\_03 & \tt 2p3 & 3627 & 199 & 330.00 & 311.35 & 59301.76 &   \cmark   \\
					\tt scene\_mp\_2p\_04 & \tt 2p4 & 1937 & 63 & 240.00 & 310.00 & 52910.41 & \\
					\tt scene\_mp\_4p\_01 & \tt 4p1 & 4173 & 407 & 320.00 & 320.00 & 75593.44 & \\
					\tt scene\_mp\_4p\_02 & \tt 4p2 & 6180 & 376 & 380.00 & 525.00 & 110844.68 &   \cmark   \\
					\tt scene\_mp\_4p\_03 & \tt 4p3 & 7495 & 632 & 400.00 & 410.00 & 98462.60 & \\
					\tt scene\_mp\_6p\_01 & \tt 6p1 & 5068 & 376 & 368.42 & 498.36 & 122027.57 & \\
					\tt scene\_mp\_6p\_02 & \tt 6p2 & 5366 & 297 & 399.57 & 440.00 & 130703.25 &   \cmark   \\
					\tt scene\_mp\_6p\_03 & \tt 6p3 & 4702 & 287 & 500.00 & 500.00 & 152452.47 & \\
					\tt scene\_sp\_cha\_01 & \tt ch1 & 2425 & 174 & 230.00 & 280.00 & 46392.01 &   \cmark   \\
					\tt scene\_sp\_cha\_02 & \tt ch2 & 3436 & 254 & 335.00 & 570.00 & 176524.94 & \\
					\tt scene\_sp\_cha\_03 & \tt ch3 & 5370 & 454 & 400.00 & 430.00 & 99961.38 & \\
					\tt scene\_sp\_cha\_04 & \tt ch4 & 6616 & 623 & 440.00 & 440.00 & 124769.37 & \\
					\tt scene\_sp\_endmaps & \tt end & 8068 & 675 & 565.00 & 770.00 & 361157.48 & \\
					\tt scene\_sp\_pol\_01 & \tt pl1 & 1055 & 63 & 323.10 & 132.93 & 12888.24 & \\
					\tt scene\_sp\_pol\_02 & \tt pl2 & 5319 & 336 & 470.00 & 515.00 & 98318.65 & \\
					\tt scene\_sp\_pol\_03 & \tt pl3 & 6828 & 558 & 420.00 & 510.00 & 127591.60 & \\
					\tt scene\_sp\_pol\_04 & \tt pl4 & 6231 & 479 & 350.00 & 340.00 & 73878.85 &   \cmark   \\
					\tt scene\_sp\_pol\_05 & \tt pl5 & 3966 & 339 & 515.00 & 394.60 & 85649.98 & \\
					\tt scene\_sp\_pol\_06 & \tt pl6 & 8320 & 679 & 470.00 & 480.00 & 157202.02 &   \cmark   \\
					\tt scene\_sp\_rus\_01 & \tt rs1 & 3125 & 276 & 330.65 & 223.97 & 33855.12 & \\
					\tt scene\_sp\_rus\_02 & \tt rs2 & 2724 & 167 & 304.60 & 307.48 & 34607.91 & \\
					\tt scene\_sp\_rus\_03 & \tt rs3 & 4878 & 372 & 450.00 & 430.00 & 70253.34 &   \cmark   \\
					\tt scene\_sp\_rus\_04 & \tt rs4 & 5628 & 360 & 338.27 & 500.00 & 104914.01 & \\
					\tt scene\_sp\_rus\_05 & \tt rs5 & 6018 & 367 & 404.41 & 418.64 & 85143.08 &   \cmark   \\
					\tt scene\_sp\_rus\_06 & \tt rs6 & 7958 & 598 & 545.00 & 455.00 & 112490.37 & \\
					\tt scene\_sp\_rus\_07 & \tt rs7 & 3599 & 206 & 460.00 & 380.00 & 86095.22 & \\
					\tt scene\_sp\_sax\_01 & \tt sx1 & 2413 & 163 & 380.00 & 485.00 & 79226.77 & \\
					\tt scene\_sp\_sax\_02 & \tt sx2 & 7038 & 368 & 403.48 & 636.57 & 118046.67 &   \cmark   \\
					\tt scene\_sp\_sax\_03 & \tt sx3 & 6708 & 368 & 510.00 & 585.00 & 153676.22 & \\
					\tt scene\_sp\_sax\_04 & \tt sx4 & 6925 & 431 & 585.00 & 675.00 & 142835.44 & \\
					\tt scene\_sp\_sax\_05 & \tt sx5 & 1954 & 65 & 445.00 & 420.00 & 86942.15 & \\
					\tt scene\_sp\_sax\_06 & \tt sx6 & 4403 & 269 & 405.00 & 465.00 & 97323.07 & \\
					\tt scene\_sp\_sax\_07 & \tt sx7 & 4151 & 348 & 310.00 & 340.00 & 69575.62 &   \cmark   \\
					\botrule
				\end{tabular*}
				
				\footnotetext{
					\emph{Legend:} $n$ is the number of map vertices (i.e., corners); $h$ is the number of holes (i.e., obstacles); $x$ and $y$ are the width and height of the map, respectively; $a$ is the map's area. The~checkmark \cmark{} means that the instance was used for tuning the parameters of our approach in~Sec.~\ref{sec:tuning} and excluded from the main experiments in Sec.~\ref{sec:eval-tea}-\ref{sec:eval-dtea}.
				}
			\end{minipage}
		\end{center}
	\end{table}
	
    The resulting maps are intricate polygons defined by thousands of vertices and dozens to hundreds of holes, typically spanning approximately $400\times400$ meters. 
    Note that the choice of the unit (meters) is arbitrary, but it should align with certain parameters, such as $d$ and $d_{\mathit{samp}}$.
    Table~\ref{tab:maps} provides detailed properties of all the maps.

\end{appendices}


\bibliography{sn-article.bib}


\end{document}